\documentclass[sn-mathphys,iicol,pdflatex]{sn-jnl}

\usepackage{xcolor}
\usepackage{subcaption}
\usepackage{multirow}
\usepackage{amsmath}
\usepackage{comment}

\jyear{2022}%

\theoremstyle{thmstyleone}%
\theoremstyle{thmstyletwo}%

\theoremstyle{thmstylethree}%

\begin{document}

\title[Article Title]{Peeling away neutron skin in ultracentral collisions of relativistic nuclei}


\author[1,2]{\fnm{Nikita} \sur{Kozyrev}}\email{kozyrev.na@phystech.edu}

\author[1,2]{\fnm{Aleksandr} \sur{Svetlichnyi}}\email{aleksandr.svetlichnyy@phystech.edu}

\author[1,2]{\fnm{Roman} \sur{Nepeivoda}}\email{nepeyvoda.rs@phystech.edu}

\author*[1,2]{\fnm{Igor} \sur{Pshenichnov}}\email{pshenich@inr.ru}

\affil[1]{\orgname{Moscow Institute of Physics and Technology}, \orgaddress{\street{Institutskiy per. 9}, \city{Dolgoprudny, Moscow Region},  \postcode{141701}, \country{Russia}}}

\affil[2]{\orgname{Institute for Nuclear Research of the Russian Academy of Sciences}, \orgaddress{\street{Prospekt 60-letiya Oktyabrya 7a}, \city{Moscow}, \postcode{117312}, \country{Russia}}}


\abstract{In heavy nuclei the ratio between local densities of neutrons and protons increases towards the nuclear periphery. The excess of neutrons is known as the neutron skin (NS) with a subtle difference ($<0.5$~fm) between the r.m.s radii of the distributions of neutrons and protons. We show that the presence of NS in $^{208}$Pb leads to extra spectator neutrons in ultracentral $^{208}$Pb--$^{208}$Pb collisions at the CERN SPS and LHC. The yields of spectator neutrons and protons were calculated within a new version of Abrasion-Ablation Monte Carlo for Colliders model (AAMCC-MST) taking into account NS and pre-equilibrium clustering of spectator matter. While the average numbers of spectator neutrons and protons in ultracentral collisions vary insignificantly, the cross sections of emission of certain numbers of spectator neutrons in events with 0,1,...5 spectator protons are changed by 50--250\%,  depending on the thickness of NS. These  cross sections are less sensitive to other parameters in calculations, and their  measurements in the ALICE experiment at the LHC will make it possible to restrict the existing variety of neutron density parameterizations in $^{208}$Pb.}

\keywords{Heavy-Ion Collisions, Nuclear Structure and Reactions}



\maketitle

\section{Introduction}

In experiments on nucleus-nucleus collisions at the Large Hadron Collider (LHC) at CERN the main attention is paid to hot and dense matter created in the domain of overlap of colliding nuclei. The largest overlap volume is achieved in very central  (ultracentral) collision events, and it is smaller in peripheral collisions.  Non-interacting nucleons outside the overlap domain retain their momenta, move forward and form excited spectator matter (prefragment). Multiple production of nuclear spectator fragments in explosive  prefragment decays is a well-known phenomenon of multifragmentation~\cite{Botvina1995} mostly investigated at lower collision energies. The study of the properties of spectator matter, in particular, its neutron and proton content is an interesting topic by itself. The n/p-ratio of free spectator nucleons has been measured in the NA49 experiment at the CERN SPS by means of forward calorimeters~\cite{Appelshauser1998}.  The ALICE experiment at the LHC is equipped with two pairs of neutron (NZDC) and proton (PZDC) Zero Degree Calorimeters (ZDC)~\cite{Puddu2007} used, in particular, for the determination of event centrality~\cite{Abelev2013} by detecting forward spectator neutrons.

In this work, we show that the detection of spectator neutrons and protons can also be used to study the neutron skin (NS) in colliding nuclei. NS is a well-known phenomenon of an  increased ratio between local  neutron and proton densities at the periphery of a heavy or a neutron-rich nucleus. The dominance of neutrons over protons at the nuclear periphery was discovered in early experiments to study nuclear structure~\cite{Barrett1977,Tanihata1985}. The radial density distributions of neutrons and protons in medium-weight and heavy nuclei are commonly characterized by the two-parameter Fermi functions:
\begin{equation}
\rho_{n,p}(r)=\frac{\rho_{0n,p}}{1+\exp\left\lfloor (r-R_{n,p})/a_{n,p}\right\rfloor}\ ,  
\label{2pF}
\end{equation}
with the following normalization to the nuclear mass number $A$:
\begin{equation}
\int d^3r (\rho_n(r)+\rho_p(r))=A\ .
\label{2pFnorm}
\end{equation}
The respective half-density radii, $R_n$ and $R_p$, and diffuseness parameters, $a_n$ and $a_p$ are taken individually for neutrons and protons, as well as the densities, $\rho_{0n}$ and $\rho_{0p}$, in the nuclear center. As known, in light neutron-rich nuclei the effect of peripheral neutron excess is caused by a weak connection between a nuclear core and a few neutrons that form a neutron halo (NH)~\cite{Hansen1995}. The neutron halo is considered also in medium-weight and heavy nuclei when $R_n = R_p$ and $a_n > a_p$ ~\cite{Trzcinska2001}. In the case of $R_n > R_p$ and $a_n = a_p$ the surface excess of neutrons is usually considered as a neutron skin~\cite{Trzcinska2001}. Due to the relations $R_n > R_p$ and $a_n > a_p$  validated for many nuclei, including $^{208}$Pb, the classification into NH and NS cases is not quite straightforward. Therefore,  we call hereafter the surface nuclear layer with the predominance of neutrons over protons simply as NS. The thickness of NS is commonly  characterized by the difference between the root-mean-square (r.m.s.) radii calculated for the point distributions of neutrons and protons: \begin{equation}
\Delta r_{np} = \left\langle r_n^2 \right\rangle^{1/2} - \left\langle r_p^2  \right\rangle^{1/2} \ .  
\end{equation}
Here $r_{n}$ and $r_{p}$ represent the distance between the center of each neutron or proton, respectively, and the center of the nucleus.  The key point in accurate determination of $\Delta r_{np}$ in $^{208}$Pb and other nuclei consists in reducing the uncertainty of $\left\langle r_n^2 \right\rangle^{1/2}$, because proton r.m.s. radii of nuclei were accurately measured~\cite{Angeli2013}.  The values of $\Delta r_{np}$ were calculated in several papers, in particular, in Refs.~\cite{Brown2000, Centelles2010, Warda2010}. Various experimental methods were used~\cite{Trzcinska2001,Tarbert2014, Adhikari2021} to measure $\left\langle r_n^2 \right\rangle^{1/2}$ and define $\Delta r_{np}$ for $^{208}$Pb. However, the existing theoretical and experimental results on the NS thickness in $^{208}$Pb and other nuclei are characterized by large  uncertainties and often contradict each other. 

The NS thickness is one of the most fundamental properties of nuclei associated with the nuclear symmetry energy and, therefore, requiring its reliable  determination.  As follows from calculations~\cite{Dobaczewski1996}, the excess of neutrons at the periphery of nuclei close to the border of stability is well above the regular increase caused by a simple excess of neutrons over protons. This is in contrast to the absence of a significant increase of the proton radius when approaching the proton-drip line reported in the same paper~\cite{Dobaczewski1996}. The relationships between the neutron-rich skin of $^{208}$Pb and the properties of neutron-star crusts have been demonstrated~\cite{Horowitz2001}. The correlations between the NS thickness in heavy nuclei, the derivative of the nuclear symmetry energy at the reduced nuclear density and the radii of moderate mass neutron stars were investigated in Ref.~\cite{Steiner2005}.  In view of the existing uncertainties of NS properties, these and other studies strongly motivate the search for new methods to restrict $\Delta r_{np}$ in heavy nuclei, in particular, in $^{208}$Pb.

It was shown~\cite{Fang2011,Fang2010,Yan2019,Aumann2017,Bertulani2019} that certain characteristics calculated for nucleus-nucleus collisions at low and relativistic energies are sensitive to the parameters of NS. The total reaction cross section, the neutron (proton) removal cross sections as well as their ratios calculated for collisions of neutron-rich $^{48}$Ca of $100A$~MeV with $^{12}$C target by means of the statistical abrasion ablation (SAA) model linearly depend on the NS thickness in $^{48}$Ca projectile~\cite{Fang2011}.  Linear correlations between the neutron removal cross sections and the NS thickness were predicted by SAA for peripheral collisions of neutron-rich Na, P, Ca and Ni nuclei of $1A$~GeV energy with $^{12}$C~\cite{Fang2010}. As shown~\cite{Yan2019}, the ratio between neutron and proton yields, as well as the tritium to $^{3}$He ratio calculated for collisions of neutron-rich $^{42-56}$Ca isotopes of $50A$~MeV with $^{40}$Ca  demonstrate strong linear correlations with $\Delta r_{np}$ in projectile nuclei. In Refs.~\cite{Aumann2017,Bertulani2019} the potential of constraining the density dependence of the symmetry energy close to saturation density through measurements of neutron-removal cross sections in high-energy nuclear collisions of 0.4 to 1 GeV/nucleon is investigated. In order to confirm all these theoretical results~\cite{Fang2011,Fang2010,Yan2019,Aumann2017,Bertulani2019} in experiments, beams of various neutron rich Na, P, Ca and Ni isotopes are necessary, while many of these isotopes are very short-lived. This creates difficulties in conducting such experiments.   

In contrast to Refs.~\cite{Fang2011,Fang2010,Yan2019,Aumann2017,Bertulani2019}, the production of secondary particles other than nucleons in $^{208}$Pb--$^{208}$Pb and p--$^{208}$Pb interactions at much higher collision energies available at the LHC were considered in Refs.~\cite{De2017,Paukkunen2015,Alvioli2019}. An impact of the neutron skin of $^{208}$Pb on the inclusive prompt photon production in $^{208}$Pb--$^{208}$Pb collisions at the LHC has been predicted~\cite{De2017}. As shown by calculations~\cite{Paukkunen2015,Alvioli2019}, the differences in spatial distributions of neutrons and protons in colliding nuclei induce small differences between the production of $W^+$ and $W^-$ bosons in peripheral p--$^{208}$Pb collisions. The authors of~\cite{De2017,Paukkunen2015,Alvioli2019} have proposed  relativistic proton-nucleus and nucleus-nucleus collisions as tools to study NS, but also pointed out the difficulties that can arise in detecting such subtle NS effects in experiments at the LHC.

In view of the above-mentioned difficulties of the experiments revealing the presence of NS in nuclei, it is worthwhile to propose measurements of other characteristics of nucleus-nucleus collisions sensitive to the parameters of NS. As shown in  Ref.~\cite{Li2020}, the charged hadron multiplicity difference between central collisions of isobar nuclei $^{96}$Ru--$^{96}$Ru and $^{96}$Zr--$^{96}$Zr at RHIC is sensitive to the neutron skin thickness. In this work we demonstrate that the calculated cross sections of emission of certain numbers of forward spectator neutrons and protons produced specifically in ultracentral collisions of relativistic $^{208}$Pb nuclei are very sensitive to the parameters of NS in $^{208}$Pb and less sensitive to other parameters in calculations. Indeed, spectator matter in ultracentral collisions is represented mostly by nucleons which are peeled away from the nuclear periphery representing NS. It is expected that in collisions of relativistic nuclei spectator nucleons are kinematically separated from participant nucleons, and the excess of neutrons with respect to protons can be detected with neutron and proton ZDCs in the ALICE experiment at the LHC~\cite{Puddu2007,Abelev2013}. 

Our suggestions~\cite{Pshenichnov2021properties, Dmitrieva2021, Kozyrev2021,Pshenichnov2022} to study ultracentral collisions of $^{208}$Pb--$^{208}$Pb at the CERN SPS and LHC complemented proposals~\cite{De2017,Paukkunen2015,Alvioli2019} to probe NS in peripheral collisions of these nuclei at the LHC. Previous studies of $^{208}$Pb--$^{208}$Pb collisions at both facilities~\cite{Appelshauser1998,Abelev2013} included detection of spectator neutrons and protons. Therefore, the effects of NS in ultracentral $^{208}$Pb--$^{208}$Pb collisions can be investigated and compared at significantly different collision energies $\sqrt{s_{\mathrm{NN}}} = 17.21$~GeV of the CERN SPS and  $\sqrt{s_{\mathrm{NN}}} = 5.02$~TeV of the LHC. The motivation to study spectator neutrons and protons specifically in most central collisions of heavy relativistic nuclei was first formulated in our papers~\cite{Pshenichnov2021properties, Dmitrieva2021, Kozyrev2021,Pshenichnov2022} based on an early version of Abrasion-Ablation Monte Carlo for Colliders (AAMCC) model. Later, the probing of neutron-skin thickness with spectator neutrons in ultracentral high-energy isobaric $^{96}$Zr--$^{96}$Zr and $^{96}$Ru--$^{96}$Ru collisions was considered independently by Liu and co-authors~\cite{Liu2022}. 

\section{Modeling spectator matter with AAMCC-MST}\label{sect_2}

A new version of the AAMCC model is used in the present work to calculate the properties of spectator matter in relativistic collisions of heavy nuclei on the event-by-event basis.  The previous AAMCC version~\cite{svetlichnyi2021using, svetlichnyi2020formation,nepeivoda2021dependence,Pshenichnov2021properties,Dmitrieva2021} has been supplemented with a pre-equilibrium decay model based on the MST-clusterisation to build the new  version of the model called AAMCC-MST~\cite{Nepeivoda2022}.  Each collision event is modelled with AAMCC-MST in several stages. Firstly, the sizes and shapes of spectator prefragments from both colliding nuclei are defined at the abrasion stage using the Glauber Monte Carlo (Glauber MC) model~\cite{Loizides2018} as described in Section~\ref{GlauberMC}. Secondly, the excitation energy of the prefragments is calculated with a hybrid approach combining the Ericson formula~\cite{Ericson1960} and ALADIN parameterization~\cite{Botvina1995}, see Section~\ref{ExcitEn}. Thirdly, at the ablation stage the minimum spanning tree (MST) clustering algorithm is applied to prefragments to define secondary clusters and their excitation energies, see Section~\ref{MST}. Finally, cluster decays are simulated via the evaporation model~\cite{Weisskopf1940}, Fermi Break-up model~\cite{Fermi1950} and Statistical Multifragmentation Model (SMM)~\cite{Bondorf1995} from Geant4 toolkit~\cite{Agostinelli2003}, see Section~\ref{PrefDecays}.

\subsection{Modeling collisions with Glauber Monte Carlo}\label{GlauberMC}

\begin{figure}[tbp!]
\centerline{\includegraphics[width=1\linewidth]{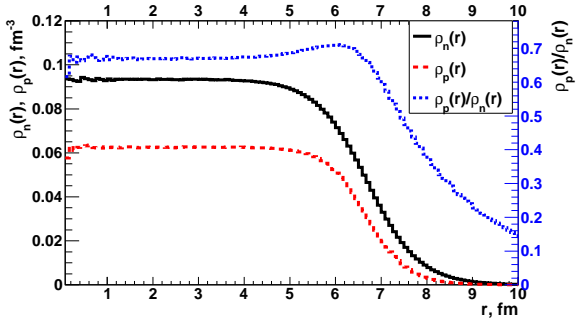}}
\caption{Radial distributions of neutron and proton densities  $\rho_{n,p}(r)$ and their ratio $\rho_p(r)/\rho_n(r)$ in  $^{208}$Pb calculated with GlauberMC profile Pbpnrw~\cite{Loizides2018} with $R_{n,p}$ and $a_{n,p}$ given in Table~\protect\ref{tab:ns_par}.}
\label{fig:208pb}
\end{figure}

The Glauber MC model~\cite{Loizides2018} is used to simulate the initial geometry of each collision event. At this first abrasion step, the positions of nucleons in colliding nuclei are sampled according to the nucleon density distributions in these nuclei, taking into account a minimum possible distance between the centers of nucleons representing the repulsive core of the nucleon-nucleon potential.  Examples of distributions of neutron and proton densities in $^{208}$Pb simulated on the basis of the two-parameter Fermi functions given by Eq.~(\ref{2pF}) are shown in Fig.~\ref{fig:208pb}. The ratio of these distributions  $\rho_p(r)/\rho_n(r)$ is also shown in Fig.~\ref{fig:208pb}. It becomes small at the nuclear periphery revealing the NS effect in $^{208}$Pb. The parameters of the Pbpnrw nuclear density profile used by default in Glauber MC for $^{208}$Pb are $R_{n}=6.69$~fm, $a_{n}=0.56$~fm and  $R_{p}=6.68$~fm, $a_{p}=0.447$~fm~\cite{Loizides2018}.

In the present work six different combinations of neutron and proton density distributions in $^{208}$Pb were implemented in AAMCC-MST. This allows us to study the sensitivity of neutron and proton content of spectator matter from ultracentral $^{208}$Pb--$^{208}$Pb collisions to the parameters of NS in $^{208}$Pb. All profiles were taken following Eqs.~(\ref{2pF}) and (\ref{2pFnorm}) with their parameters listed in Table~\ref{tab:ns_par}. These profiles were grouped  into three pairs to compare AAMCC-MST results pairwise. The density profiles in the first pair, denoted as NFB-8  and  NL1, were taken from Ref.~\cite{Centelles2010}. They represent calculated density distributions which characterize, respectively, the most thin ($\Delta r_{np}=0.115$~fm) and thick ($\Delta r_{np}=0.321$~fm) NS among all the nuclear structure models of $^{208}$Pb mentioned in Ref.~\cite{Centelles2010}. The parameterizations Pbpnrw and  PREX in the second pair are based on the results of Crystal Ball at MAMI and A2 collaboration~\cite{Tarbert2014} and PREX~\cite{Adhikari2021} collaboration, respectively. These collaborations conducted measurements for $^{208}$Pb by two different techniques and used different theoretical models to fit experimental results. In Ref.~\cite{Tarbert2014} an interpolated fit of a theoretical model to the measured cross sections provided $R_n=6.70\pm 0.03$~fm and $a_n=0.55\pm 0.01\mathrm{(stat.)}-0.03+0.02\mathrm{(sys.)}$~fm. Following this result, the parameterization Pbpnrw with $R_n=6.69$~fm and $a_n=0.56$~fm was adopted in Glauber MC as a default option for $^{208}$Pb~\cite{Loizides2018}. In Ref.~\cite{Adhikari2021} $\Delta r_{np}=0.283\pm 0.071 $~fm was extracted from measurements of parity violation in elastic scattering of polarised electrons from $^{208}$Pb, but the corresponding $R_n$ and $a_n$ were not reported. As known~\cite{Warda2010}, even with a well-established parametrization of proton density  in $^{208}$Pb with $R_p=6.68$~fm and  $a_p=0.447$~fm~\cite{Tarbert2014,Loizides2018}, many combinations of $R_n$ and $a_n$ lead to the same $\Delta r_{np}$. 

In order to find $R_n$ and $a_n$ which reasonably reproduce $\Delta r_{np}=0.283$~fm reported by the PREX collaboration~\cite{Adhikari2021}, the average value of $R_n=6.810$~fm from several versions of the models considered in Ref.~\cite{Centelles2010} was taken. Then, the diffuseness parameter $a_n=0.60$~fm for neutrons was calculated from the following relations~\cite{Warda2010}: 
\begin{equation}
\left\langle r_{p}^2 \right\rangle = \frac{3}{5} R_p^2\left(1+\frac{7}{3}\frac{\pi^2a_p^2}{R_p^2}\right), \label{Eq:r_p}
\end{equation}
\begin{equation}
\left\langle a_{n}^2 \right\rangle = \frac{5}{7\pi^2}\left(\Delta r_{np} + \sqrt{\left\langle r_{p}^2 \right\rangle}\right)^2 - \frac{3}{7}\frac{R_n^2}{\pi^2}. \label{Eq:a_n}
\end{equation}

The values of $R_n=6.810$~fm, $a_n=0.60$~fm, $R_p=6.68$~fm and $a_p=0.447$~fm thus represent the nuclear density profile of $^{208}$Pb denoted as PREX, see Table~\ref{tab:ns_par}. The Pbpnrw  and  PREX profiles also represent thin ($\Delta r_{np}=0.150$~fm) and thick ($\Delta r_{np}=0.283$~fm) NS, respectively. 

With the aim to evaluate the uncertainty of selecting $R_n$ and $a_n$ to reproduce $\Delta r_{np}=0.283$~fm reported by the PREX collaboration, two additional parameterizations denoted as PREX1 and PREX2 were introduced as a third pair in Table~\ref{tab:ns_par}. These parameterizations provide the same $\Delta r_{np}=0.283$ fm, but with the minimum and maximum values of $R_n$ corresponding, respectively, to NFB-8  and  NL1 profiles~\cite{Centelles2010}. Following  Ref.~\cite{Trzcinska2001}, PREX1 profile with its parameters $R_n=R_p$ and $a_n>a_p$ can be considered as a pure case of a neutron halo (NH) in $^{208}$Pb. Other profiles, excluding NFB-8, have 
$R_n>R_p$ and $a_n>a_p$ and thus should be considered as nuclear density profiles with NS.
\begin{table*}[tp]
\centering
\caption{Half-density radii $R_{n,p}$, diffuseness parameters $a_{n,p}$ and the corresponding neutron skin thickness $\Delta r_{np}$ for six different parameterizations of neutron and proton densities in $^{208}$Pb used in this work in AAMCC-MST calculations. The values of $b_{max}$ for the impact parameter interval $[0,b_{max}]$ corresponding to 0--5\% centrality at $\sqrt{s_{\mathrm{NN}}} = 17.21$~GeV and $5.02$~TeV, respectively, are also given.}
\begin{tabular}{|c|ccccc|cc|}
\hline
 & $R_n$, & $a_n$, & $R_p$, & $a_p$, & $\Delta r_{np}$, & \multicolumn{2}{c|}{$b_{max}$,} \\
 &  fm   &  fm   &  fm   & fm    &  fm & \multicolumn{2}{c|}{fm} \\
 &       &       &       &       &     & 17.21 GeV    & 5.02 TeV    \\
\hline
NFB-8 & $6.679$ & $0.546$ &  $6.683$ & $0.451$ & $0.115$ & $3.32$ & $3.45$\\
NL1 & $6.940$ & $0.587$ & $6.718$  & $0.463$ & $0.321$ & $3.45$ & $3.57$ \\
\hline
Pbpnrw & $6.69$ & $0.56$ & $6.68$  & $0.447$ & $0.15$ & $3.37$ & $3.49$ \\
PREX & $6.81$ & $0.60$ & $6.68$  & $0.447$ & $0.283$ & $3.42$ & $3.54$ \\
\hline
PREX1 & $6.68$ & $0.66$ & $6.68$  & $0.447$ & $0.283$ & $3.46$ & $3.59$ \\
PREX2 & $6.94$ & $0.53$ & $6.68$  & $0.447$ & $0.283$ & $3.38$ & $3.50$ \\
\hline
\end{tabular}
\label{tab:ns_par}
\end{table*}

In contrast to the uncertainties of the parameters of neutron distributions in $^{208}$Pb, the profiles Pbpnrw  and  PREX, as well as PREX1 and PREX2, are based on the well-established parameters for proton density distributions: $R_{p}=6.68$~fm and $a_{p}=0.447$~fm, see Ref.~\cite{Loizides2018} for details. Therefore, these values of $R_{p}$ and $a_{p}$ were used in the calculations with these four density profiles.

In modelling of each event the centers of the colliding nuclei are placed initially at ($\pm \frac{b}{2}$, 0, 0), where $b$ is the impact parameter of the collision. The impact parameter $b$ is defined as the distance of the closest approach between the centers of colliding nuclei. It is sampled following the distribution of the number of events per unit interval of $b$ taken as $\frac{dN}{db}\propto b$. Ultracentral $^{208}$Pb--$^{208}$Pb collisions modelled in this work were defined as events of 0--5\% centrality corresponding to the impact parameter interval of $[0,b_{max}]$. Because of slightly different radial extensions of nuclear density in each of the six parametrerizations, $b_{max}$ values corresponding to 0--5\% centrality are also different, and they are listed in Table~\ref{tab:ns_par} separately for  $^{208}$Pb--$^{208}$Pb collisions  at $\sqrt{s_{\mathrm{NN}}} = 17.21$~GeV and $5.02$~TeV. As in other implementations of the Glauber model, it is assumed that in the course of each collision all nucleons move strictly parallel to the beam axis $z$  without changing their positions in the transverse ($x,y$) plane. 

In Glauber MC model there are two options to calculate the probability of a nucleon-nucleon collision $P(b_\mathrm{NN})$ as a function of NN impact parameter $b_\mathrm{NN}$. In the first option  nucleons are considered as hard spheres with the diameter $D=\sqrt{\sigma^{\mathrm{NN}}_{inel}/\pi}$. Two nucleons from colliding nuclei are considered interacting, and thus termed participants, if the distance between their centers in the transverse  ($x,y$) plane is less than $D$. With this default  option:
\begin{equation}
P(b_\mathrm{NN}) = \Theta(D - b_\mathrm{NN}) \ . 
\label{eq:p_b_default}
\end{equation}
The fluctuations of nucleon shape or parton degrees of freedom are taken into account with another option of GlauberMC~\cite{Loizides2016} to calculate the NN collision probability as
\begin{equation}
    P(b_\mathrm{NN}) = \Gamma \Big(1/w, \frac{b_\mathrm{NN}^2}{D^2w} \Big)\Big/\Gamma (1/w) \ .
\label{eq:p_b}
\end{equation}
Here $\Gamma$ is the gamma function and $w$ is a parameter which controls the transition from the hard-sphere ($w \rightarrow 0$) to the Gaussian ($w \rightarrow 1$) representation of $P(b_\mathrm{NN})$. As pointed out in Ref.~\cite{Loizides2018}, an intermediate value of $w$ = 0.4 reproduces well the values of the total and elastic $pp$ cross sections measured at the LHC. As also shown~\cite{Loizides2018}, by using $w$ = 0.4 in Eq.~(\ref{eq:p_b}) the number of binary nucleon-nucleon collisions $N_{coll}$ is reduced  up to 10\% with respect to calculations with hard-sphere option Eq.~(\ref{eq:p_b_default}) in modelling peripheral $^{208}$Pb--$^{208}$Pb  collisions at $\sqrt{s_{\mathrm{NN}}} = 5.02$~TeV. However, a  smaller reduction of $N_{coll}$ by $\sim 1$\%  is estimated for central (0--5\%) events. 

The dependence of $\sigma^{\mathrm{NN}}_{inel}$ on the square of center-of-mass energy $s$ in nucleon-nucleon collisions is parameterized by $\sigma^{\mathrm{NN}}_{inel}(s) = {\cal A} + {\cal B}\times \ln^2(s)$, where the parameters ${\cal A}$ and ${\cal B}$ are extracted from experimental data, see Ref.~\cite{Loizides2018} for details. While each nucleon can interact several times with nucleons of the collision partner, same inelastic nucleon-nucleon cross section is assumed for the first and  for all subsequent collisions despite of nucleon energy loss. This assumption is based on the weak dependence of $\sigma^{\mathrm{NN}}_{inel}$ on $\mathrm{NN}$ collision energy typical for high energies.  It is assumed that all other nucleons that are not participants form two excited spectator prefragments representing the remnants of colliding nuclei.

\subsection{Calculation of prefragment excitation energy}\label{ExcitEn}

Each of the two prefragments is produced by removing $a$ participating nucleons from the respective initial nucleus with the mass number $A$ by the end of the abrasion stage. Since an individual nucleon can interact with one or more nucleons from another nucleus, the number of participants in each of the two nuclei usually differs. Such fluctuations are properly reflected by Monte Carlo modeling, and the prefragment mass number is calculated individually on each side as $A_{pf}=A-a$. Two different methods to calculate the prefragment excitation energy $E^\star$ are employed in AAMCC-MST depending on the relative mass of the prefragment $\alpha_{\rm pf} = A_{\rm pf}/A$.   

In the case of a peripheral collision event when very few nucleons are removed, $a\ll A$, it is assumed that the properties of a nuclear residue remain similar to those of the initial nucleus. As discussed~\cite{Gaimard1991}, this justifies the calculation of the total prefragment excitation energy by considering a set of nucleon holes created in the initial nucleus. With this method the distribution of the excitation energy of a prefragment is computed from the density of states of the nucleus $A$ with $a$ nucleon holes. Following the calculations on the basis of Ericson formula~\cite{Ericson1960}, this results in the mean excitation energy per removed nucleon in events with $a$ removed nucleons~\cite{Scheidenberger2004}: 
\begin{equation}
   \frac{\langle E^\star\rangle}{a}  = E_{\rm max} \frac{a}{a+1}\ . 
    \label{Eq:E_a}
\end{equation}
Here $E_{\rm max} = 40$~MeV sets the maximum energy of a single hole state in the initial heavy nucleus~\cite{Scheidenberger2004}.  From Eq.~(\ref{Eq:E_a}) the average excitation energy per prefragment nucleon $\langle \epsilon^* \rangle=\langle E^*/A_{\rm pf}\rangle $ can be obtained:
\begin{equation}
\langle \epsilon^\star \rangle = E_{\rm max}\frac{(1-\alpha_{\rm pf})^2}{\alpha_{\rm pf}\big(1-\alpha_{\rm pf}+A^{-1}\big)}\  \ ,
\label{eq:eric}
\end{equation}
with $\alpha_{\rm pf}=A_{pf}/A$ defined as a relative mass of the prefragment. In central collisions the numbers of removed nucleons are comparable to their total numbers in the initial nuclei, $a\sim A$, and essential changes of the both nuclear cores are expected. In this case the validity of Eq.~(\ref{eq:eric}) becomes questionable, and it is replaced by a phenomenological dependence:
\begin{equation}
   \langle \epsilon^* \rangle= \epsilon_{\rm max} \sqrt{1 - \alpha_{\rm pf}} \  \  .
    \label{eq:aladin}
\end{equation}
This relation between $\langle \epsilon^* \rangle$ and $\alpha_{\rm pf}$ is similar to the one used in Ref.~\cite{Botvina1995}, but without its linear term. The value of $\epsilon_{\rm max} = 11.5$~MeV is used in the present work as a model parameter. A dispersion given by 
\begin{equation}
    \sigma = \sigma_0 (1 + b_0 \cdot(1 - \alpha_{\rm pf})) \label{eq:sigma}
\end{equation}
around the average values of $\alpha_{\rm pf}$ was assumed with 
$\sigma_0 = 0.005$ and $b_0 = 2$ taken as free parameters. 

It was found that the calculated average multiplicities of spectator neutrons  $\langle N_n \rangle$ and protons $\langle N_p \rangle$ as functions of the collision impact parameter $b$ are mostly defined by the dependence of  $\langle \epsilon^* \rangle$ on $\alpha_{\rm pf}$, rather than by the dispersion $\sigma$. Due to the overestimation of $E^\star$ in peripheral collisions, $\langle N_n \rangle$ and $\langle N_p \rangle$ are also overestimated in calculations based on Eq.~(\ref{eq:aladin}) with respect to the neutron multiplicities measured in peripheral $^{208}$Pb--$^{208}$Pb collisions~\cite{CollaborationALICE2020}. In contrast, calculations based on the Ericson formula, Eq.~(\ref{Eq:E_a}), overestimate the multiplicities of spectator neutrons and protons in semi-central and central $^{208}$Pb--$^{208}$Pb collisions when  compared with the data~\cite{CollaborationALICE2020}. In order to resolve this issue, a hybrid parameterization of the prefragment excitation energy was introduced in AAMCC-MST in Ref.~\cite{Nepeivoda2022} to improve the description of ALICE data~\cite{CollaborationALICE2020}. In peripheral events for prefragments with $\alpha_{\rm pf} > \alpha_{\mathrm sw}$ the excitation energy is calculated on the basis of Ericson formula, while for events with low relative mass of prefragments, $\alpha_{\rm pf} < \alpha_{\mathrm sw}$, their excitation energy is calculated according to the ALADIN parameterization. Then, the dependence (\ref{eq:eric}) is switched to (\ref{eq:aladin}) at the value of $\alpha_{\rm sw}$ determined from the condition of continuity of the expressions (\ref{eq:eric}) and (\ref{eq:aladin}) used at low and high $\epsilon^\star$, respectively. Neglecting $1/A$ for heavy nuclei and equating the expressions (\ref{eq:eric}) and (\ref{eq:aladin}) gives
\begin{equation}
\alpha_{\rm sw}=\frac{\sqrt{1+4\kappa}-1}{2\kappa},\ \ \kappa=\frac{\epsilon_{\rm max}^2}{E_{\rm max}^2}\ . 
\label{eq:alphaswitch}
\end{equation}
With the parameters used in the present work for $^{208}$Pb projectiles Eq.~(\ref{eq:alphaswitch}) provides $\alpha_{\rm sw}=0.9287$. This means that for central $^{208}$Pb--$^{208}$Pb collisions, which are of the main interest in the present work, the method based on the ALADIN parameterization, Eq.~(\ref{eq:aladin}), is applied.

\subsection{Preequilibrium clustering of spectator matter}\label{MST}

\begin{figure}[htbp!]
\begin{minipage}[h]{1\linewidth}
\center{\includegraphics[width=1.07\linewidth]{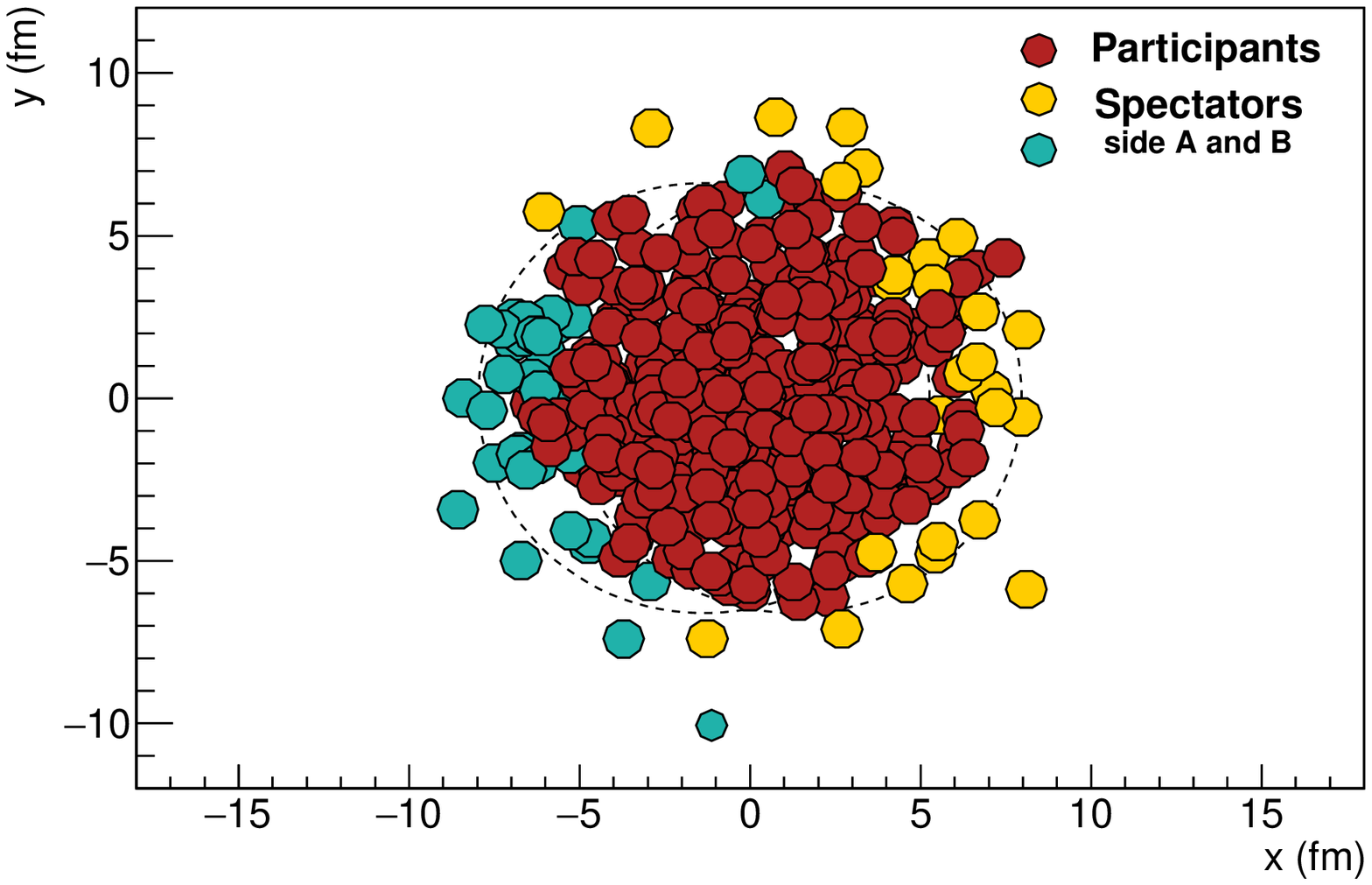}} \\
\end{minipage}
\vfill
\begin{minipage}[htbp!]{1\linewidth}
\center{\includegraphics[width=1.07\linewidth]{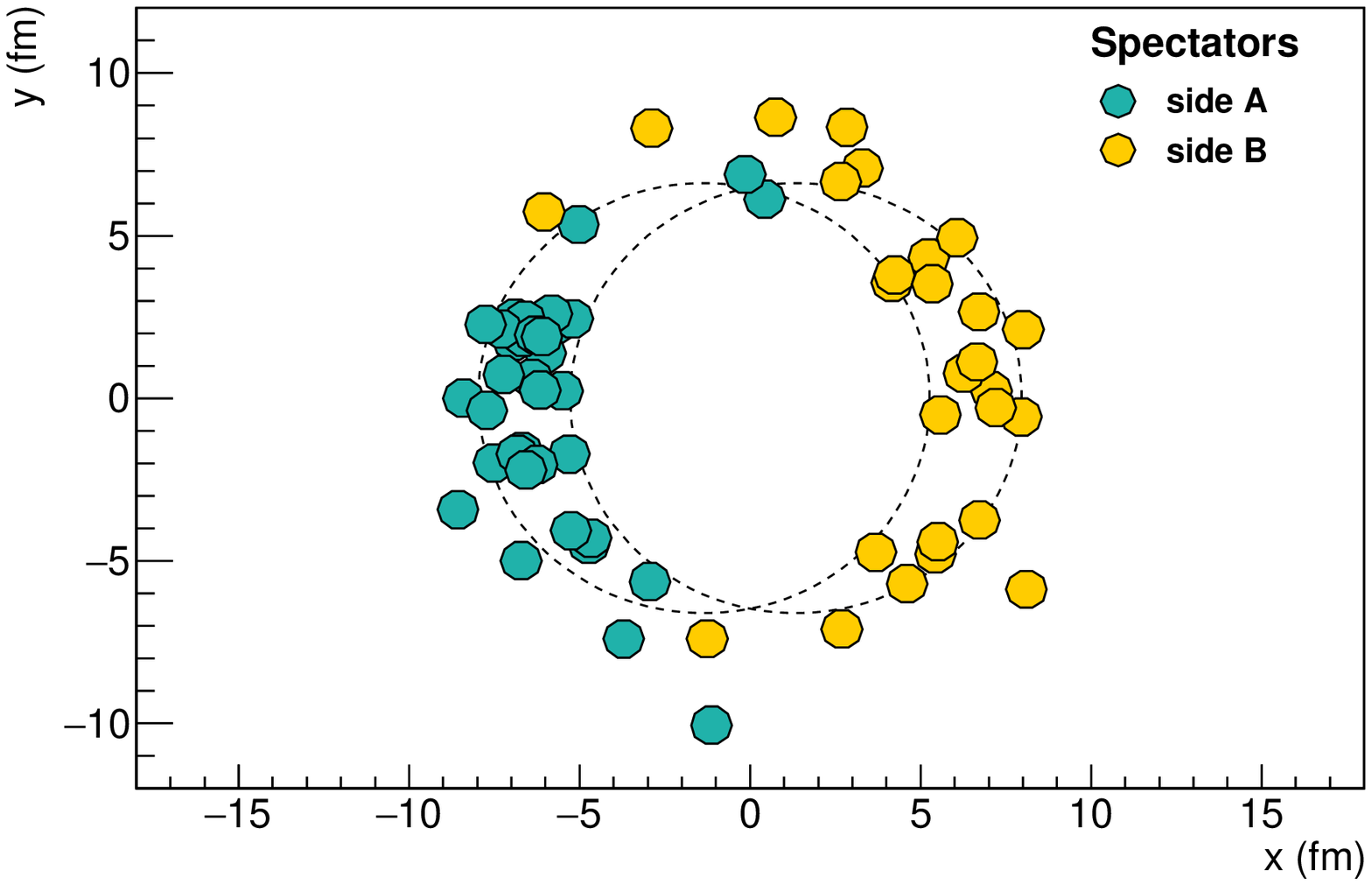}} \\
\end{minipage}
\hfill
\caption{Transverse projection of a central $^{208}$Pb$-^{208}$Pb collision at $\sqrt{s_{NN}}=17.2$ GeV (top). Spectator nucleons from each of the colliding nuclei in the same event are shown separately in the bottom panel.}
\label{fig:coll_geom}
\end{figure}

To the best of our knowledge, in all versions of the abrasion-ablation model spectator matter (a prefragment) is considered as an entire nuclear system with intense interactions between constituent nucleons. Therefore, it is  assumed that the  thermodynamic equilibrium in the entire prefragment is achieved by the time of its decay. However, the validity of this assumption can be questioned in the case of central nucleus-nucleus collisions, in which the prefragment has a shape of a narrow crescent, see Fig.~\ref{fig:coll_geom}. It can be expected that in such collisions the hot prefragment disintegrates into several clusters already at the preequilibrium stage and later the thermodynamic equilibrium is established locally in each cluster. Similarly to the assumptions adopted in Ref.~\cite{Botvina2022}, an excited nuclear system undergo a subdivision into primary equilibrated nucleon clusters, and final state nuclei are produced after the decay of these excited clusters.    

In order to account for preequilibrium disintegration of prefragments we have developed an algorithm of clusterization based on the construction of a minimum spanning tree (MST)~\cite{Prim1957}. In this algorithm, groups of nucleons of spectator matter are identified as clusters on the basis of their proximity in three-dimensional coordinate space.

First, it is assumed that all nucleons represent the vertices of a complete weighted undirected graph. The edge weights of the graph are equal to the modulus of the distance between the corresponding vertex nucleons. Second, a tree without breaking connectivity with the minimum possible sum of all the edge weights is constructed by means of Kruskal's algorithm.  Third, the critical distance cut is applied to search for inconsistent edges and subdivide the MST tree. This is made by removing edges with weights greater than some critical distance $d$ considered as a free parameter of the model. Finally, the depth-first search is used to define clusters. 

The application of the MST algorithm to large prefragments resulting from peripheral collisions should not lead to their separation into many clusters at the preequilibrium stage. Therefore, it is reasonable to set $d$ significantly longer than the average distance between the centers of nucleons in nuclear matter at the normal nuclear density. In particular, in modelling peripheral collisions assuming $d_0=2.7$~fm, prefragments remained  sufficiently stable. At the same time, prefragements with specific shapes produced in central collisions were dissociated into clusters in calculations with $d_0=2.7$~fm.

One can expect a higher average multiplicity of nucleons and nucleon clusters estimated by the MST algorithm for dilute nuclear matter with reduced density $\rho<\rho_0$ with respect to prefragments of the normal density $\rho_0$.  
The AAMCC-MST model is based on the Glauber theory, and estimating the prefragment density   at the preequilibrium stage is not a simple task. During a short collision time any changes in the spatial coordinates of nucleons due to the radial expansion of the spectator matter in the plane transverse to the beam axis are neglected in the Glauber theory. Alternatively, the lower limit for the prefragment density at the preequilibrium stage can be estimated from the average prefragment density $\langle\rho\rangle$ at a later stage, after reaching the thermodynamic equilibrium.

As estimated from experimental data~\cite{Viola2004,De2006}, at high $\epsilon^\star$, $\langle\rho\rangle$ drops significantly, and at $\epsilon^\star\ge 5$~MeV it reaches a plateau value of $\langle\rho\rangle\sim (1/3-1/2)\rho_0$. Similarly, in the Statistical Multifragmentation Model (SMM)~\cite{Bondorf1995} it is assumed that at the moment of an explosive decay of a hot termalized nuclear system with $\epsilon^\star\ge 3-4$~MeV into fragments, its average density is also lower than $\rho_0$: $\langle\rho\rangle\sim (1/4-1/3)\rho_0$. 

The dependence of $\langle\rho(\epsilon^\star)/\rho_0\rangle$ on the excitation energy $\epsilon^\star$ at the time of  preequilibrium clusterization is postulated in the AAMCC-MST model taking into account the above mentioned results~\cite{Viola2004,De2006}. It is assumed that $\langle\rho(\epsilon^\star)/\rho_0\rangle=1$ at $\epsilon^\star\leq \epsilon_0$. At higher excitation energies $\epsilon^\star>\epsilon_0$ a phenomenological approximation of $\langle\rho(\epsilon^\star)/\rho_0\rangle$ by a power function is introduced to account for the expansion:
\begin{equation}
\langle\rho(\epsilon^\star)/\rho_0\rangle = (\epsilon^\star/\epsilon_0)^{\gamma}\ .
\label{eq:powerfit}
\end{equation}
Here $\gamma = -1.02 \pm 0.10$ and $\epsilon_0 = 2.17 \pm 0.23$~MeV are the parameters which ensure the continuity of $\langle\rho(\epsilon^\star)/\rho_0\rangle$ below and above $\epsilon_0$.

Following Eq.~(\ref{eq:powerfit}) a phenomenological dependence of the MST clustering parameter $d$ on the excitation energy at the preequilibrium stage is postulated in the AAMCC-MST model. It is expected that due to more intense motion of intracluster nucleons and outward pressure associated with the increase of $\epsilon^\star$, the internal density of clusters decreases on the way to the thermal equilibrium. However, due to a short duration of the preequilibrium stage, there is no significant global expansion of the cluster system. Therefore, with an increase in the excitation energy of the system, the distances between the cluster boundaries also decrease because they correlate on average with $d$. Since the exact dependence of $d$ on the internal density of clusters is unknown, a phenomenological dependence of $d\propto\rho^{1/3}(\epsilon^\star)$ is assumed. Together with Eq.~(\ref{eq:powerfit}) it finally gives:
\begin{equation}
d = 
\begin{cases}
d_0, \  \epsilon^\star < \epsilon_0 \\
d_0  \  (\epsilon^\star/\epsilon_0)^{\gamma/3}, \  \epsilon^\star > \epsilon_0
\end{cases}
\text{, }
\label{eq:clustdist}
\end{equation}
where $d_0 = 2.7$~fm is the clustering parameter at the normal nuclear density $\rho_0$. The decrease of $d$ with the increase of $\epsilon^\star$ reflects a decrease of the connectivity of the prefragment system with the growth of $\epsilon^\star$. It is assumed that the excitation energy $E^\star$ of the spectator prefragment is distributed between the resulting clusters proportionally to the number of nucleons they contain.

\subsection{Modeling decays of clusters}\label{PrefDecays}

Statistical models of the Geant4~\cite{Allison2016} toolkit are used in AAMCC-MST to simulate decays of clusters formed at the preequilibrium stage. For clusters with $\epsilon^\star\leq 4$~MeV, the Weisskopf-Ewing model~\cite{Weisskopf1937} is used to simulate sequential evaporation of nucleons and light nuclei ($^{2}$H, $^{3}$H, $^{3}$He, $^{4}$He).  The Statistical Multifragmentation Model (SMM)~\cite{Bondorf1995} of  Geant4~\cite{Allison2016} is employed at $\epsilon^\star> 4$~MeV. The Fermi Break-Up model~\cite{Fermi1950} from the same toolkit is used to simulate decays of excited light nuclei up to $^{18}$O.

Specific versions of Geant4 decay models were employed on the basis of the correspondence of their results to Fortran versions of the respective models~\cite{Bondorf1995}. Namely, the SMM and evaporation models were taken from the Geant4 of version 10.4, while the Fermi Break-up model was adopted from the Geant4 of version 9.6.

\section{Results and discussion}

\subsection{Average numbers of spectator neutrons and protons}\label{Subsection_3_1}

Six different parameterizations of neutron and proton densities in $^{208}$Pb listed in Table~\ref{tab:ns_par} were used in AAMCC-MST to calculate the average numbers of spectator neutrons $\left\langle N_n \right\rangle$ and protons $\left\langle N_p \right\rangle$ in central $^{208}$Pb--$^{208}$Pb collisions at the CERN SPS. The data of the NA49 collaboration~\cite{Appelshauser1998} obtained for $^{208}$Pb--$^{208}$Pb collisions with $\langle b \rangle=2$~fm  at the CERN SPS at $\sqrt{s_{\mathrm{NN}}} = 17.21$~GeV  make it  possible to validate the AAMCC-MST results. As reported~\cite{Appelshauser1998}, forward-going spectator matter in central $^{208}$Pb--$^{208}$Pb  collisions consists on average of 9 neutrons, 7 protons, and 0.5 deuterons. The values of  $\left\langle N_n \right\rangle$, $\left\langle N_p \right\rangle$ and $\left\langle  N_d \right\rangle$ calculated with AAMCC-MST for $^{208}$Pb--$^{208}$Pb  collisions at $b$ = 2~fm are compared with the data~\cite{Appelshauser1998} in Table~\ref{tab:na49}. The values of $\left\langle N_p \right\rangle$ calculated with all nuclear density parameterizations agree with the measured $\left\langle N_p \right\rangle$ within the experimental uncertainties.  This indicates that the parameters of the proton density distributions used in AAMCC-MST calculations are realistic. However, the average numbers of spectator neutrons $\left\langle N_n \right\rangle$ are overestimated with all NS options.  One can note that the average $\langle b \rangle=2$~fm for central events was reported in Ref.~\cite{Appelshauser1998} rather than a specific interval of $b$. The observed disagreement between the measurements and calculations can be potentially caused by assuming the specific value of $b=2$~fm  in calculations with AAMCC-MST.  Nevertheless, an obvious correlation between $\left\langle N_n \right\rangle$ and $\Delta r_{np}$ is seen in Table~\ref{tab:na49}. As expected, slightly more neutrons are emitted in modelling with a thicker NS represented by NL1, PREX, PREX1 and PREX2 options. The calculated average numbers of deuterons are overestimated in comparison to the measured $\left\langle  N_d \right\rangle$, but some variations of deuteron yields are also seen with respect to the NS thickness. Namely, the highest $\left\langle N_d \right\rangle$ of 0.87 is obtained with the thinnest NS in NFB-8, while the lowest $\left\langle N_d \right\rangle$ of 0.77 is associated with the most thick NS in NL1. Due to the equal presence of neutrons and protons in deuterons, neutron-rich spectator matter is less favorable for the formation of deuterons.
\begin{table}[tbp!]
\centering
\caption{Average numbers of spectator neutrons $\left\langle N_n \right\rangle$, protons $\left\langle N_p \right\rangle$ and deuterons $\left\langle N_d \right\rangle$ in ultracentral ($b = 2$ fm) $^{208}$Pb--$^{208}$Pb collisions at $\sqrt{s_{\mathrm{NN}}} = 17.21$~GeV measured by NA49 collaboration~\cite{Appelshauser1998} and calculated with AAMCC-MST with different nuclear density profiles.}
\begin{tabular}{|c|c|ccc|}
\hline
   & $\Delta r_{np}$, fm  & $\left\langle N_n \right\rangle$ & $\left\langle N_p  \right\rangle$ & $\left\langle  N_d \right\rangle$ \\
\hline
 NFB-8 & 0.115 & 12.45 & 6.82 & 0.87 \\
  NL1  & 0.321 & 13.88 & 6.44 & 0.77 \\
 \hline
 Pbpnrw & 0.15 & 12.73 & 6.74 & 0.85 \\
 PREX & 0.283 & 13.59 & 6.42 & 0.80 \\
 \hline
 PREX1 & 0.283 & 13.79 & 6.49 & 0.78 \\
 PREX2 & 0.283 & 13.41 & 6.32 & 0.82 \\ 
 \hline
NA49  &  & $9.0 \pm 1.8$  &  $7.0 \pm 1.4$  &  0.5  \\
\hline
\end{tabular}
\label{tab:na49}
\end{table}
 
The values of $\left\langle N_n \right\rangle$, $\left\langle N_p \right\rangle$ were calculated with AAMCC-MST in Ref.~\cite{Nepeivoda2022} as functions of $b$ and were compared with preliminary ALICE data~\cite{CollaborationALICE2020} for $^{208}$Pb--$^{208}$Pb collisions at $\sqrt{s_{\mathrm{NN}}} = 5.02$~TeV. The  value of $\left\langle N_p \right\rangle$ calculated at $b=2$~fm was found in good agreement with measured $\left\langle N_p \right\rangle\sim 3$. At the same time, calculated $\left\langle N_n \right\rangle$ underestimated the data for central collisions at $\sqrt{s_{\mathrm{NN}}} = 5.02$~TeV, in contrast to $\left\langle N_n \right\rangle$ calculated in the present work at  $\sqrt{s_{\mathrm{NN}}} = 17.21$~GeV.

\subsection{NFB-8 versus NL1 profile and  dependence on collision energy}\label{Subsection_3_2}

The dependence of the yields of spectator nucleons in ultracentral $^{208}$Pb--$^{208}$Pb events on collision energy can be understood from the comparison of AAMCC-MST results obtained at $\sqrt{s_{\mathrm{NN}}} = 17.21$~GeV and at 5.02~TeV for the same  centrality of 0--5\%.

\begin{table}[htbp]
\centering
\caption{Average numbers of spectator neutrons $\left\langle N_n \right\rangle$ and protons $\left\langle N_p \right\rangle$ in ultracentral  $^{208}$Pb--$^{208}$Pb collisions for 0--5\% centrality at $\sqrt{s_{\mathrm{NN}}}=17.21$~GeV (SPS) and $\sqrt{s_{\mathrm{NN}}}=5.02$~TeV (LHC) calculated with NFB-8 and NL1 density profiles.}
\begin{tabular}{|p{0.08\linewidth}|p{0.08\linewidth}p{0.08\linewidth}|p{0.08\linewidth}p{0.08\linewidth}|} 
\hline
 & \multicolumn{2}{c|}{NFB-8} & \multicolumn{2}{c|}{NL1} \\ 
& $\left\langle N_n \right\rangle$ & $\left\langle N_p \right\rangle$ & $\left\langle N_n \right\rangle$ & $\left\langle N_p \right\rangle$  \\ 
\hline
SPS & 14.38 & 7.81 & 16.24 & 7.72  \\ 
\hline
LHC & 8.67 & 4.60  & 9.92  & 4.38   \\
\hline
\end{tabular}
\label{tab:nav_en}
\end{table}

As seen from Table~\ref{tab:nav_en}, $\left\langle N_n \right\rangle$ and $\left\langle N_p \right\rangle$ calculated with NFB-8, as well as with NL1 profiles,  noticeably decrease with the increase of collision energy. Such a depletion of spectator matter in collisions at the LHC is explained by the increase of the nucleon-nucleon cross section $\sigma^{\mathrm{NN}}_{inel}$ as discussed in Section~\ref{sect_2}. Because of this increase, the number of NN-collisions is larger and results in the increase of the number of participant nucleons at high energy. Consequently, the number of remaining spectator nucleons is reduced at the LHC.

\begin{figure}[tbp!]
\begin{subfigure}[b]{1.0\linewidth}
\centerline{\includegraphics[width=1.0\linewidth]{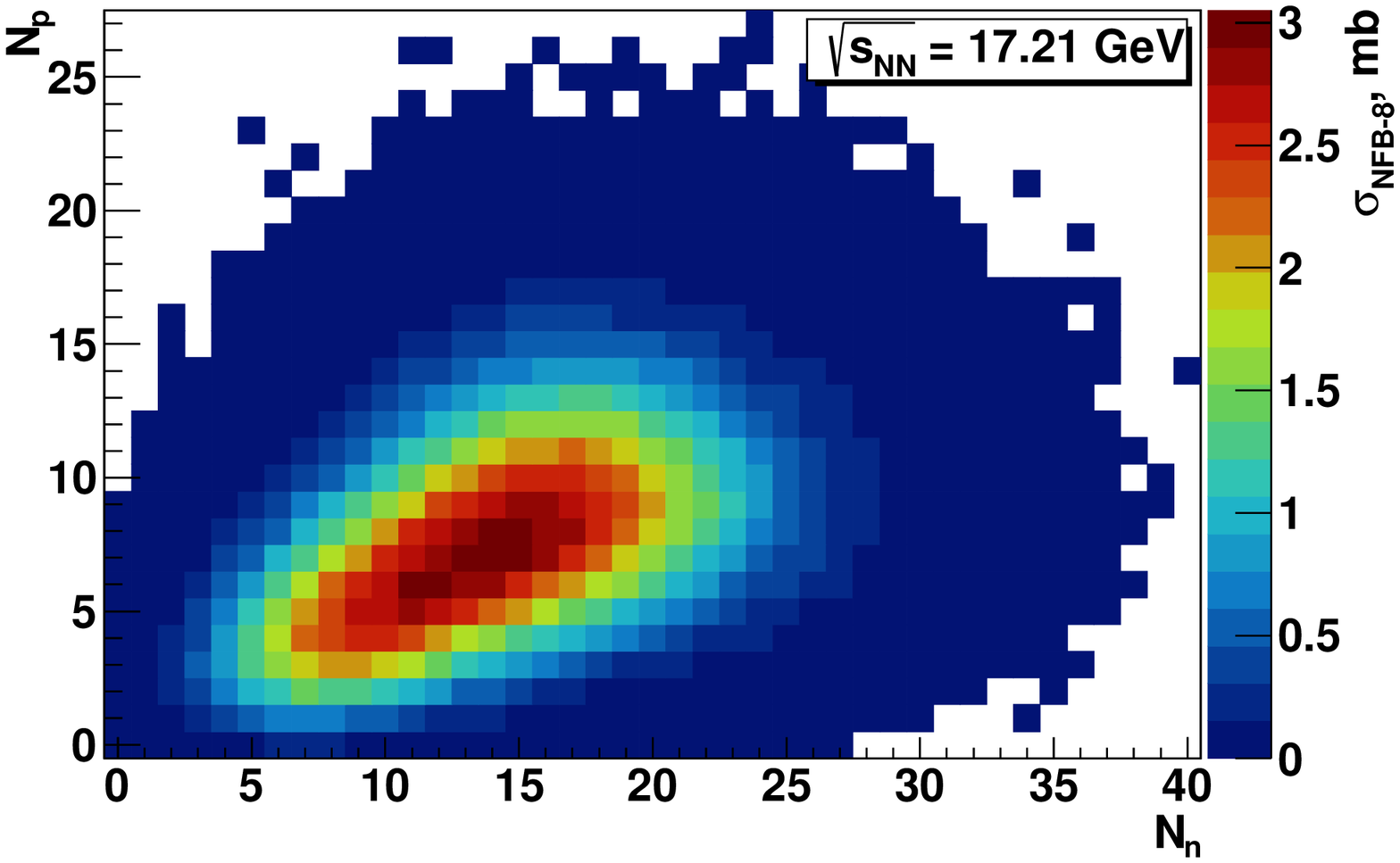}}
\end{subfigure}
\begin{subfigure}[b]{1.0\linewidth}
\centerline{\includegraphics[width=1.0\linewidth]{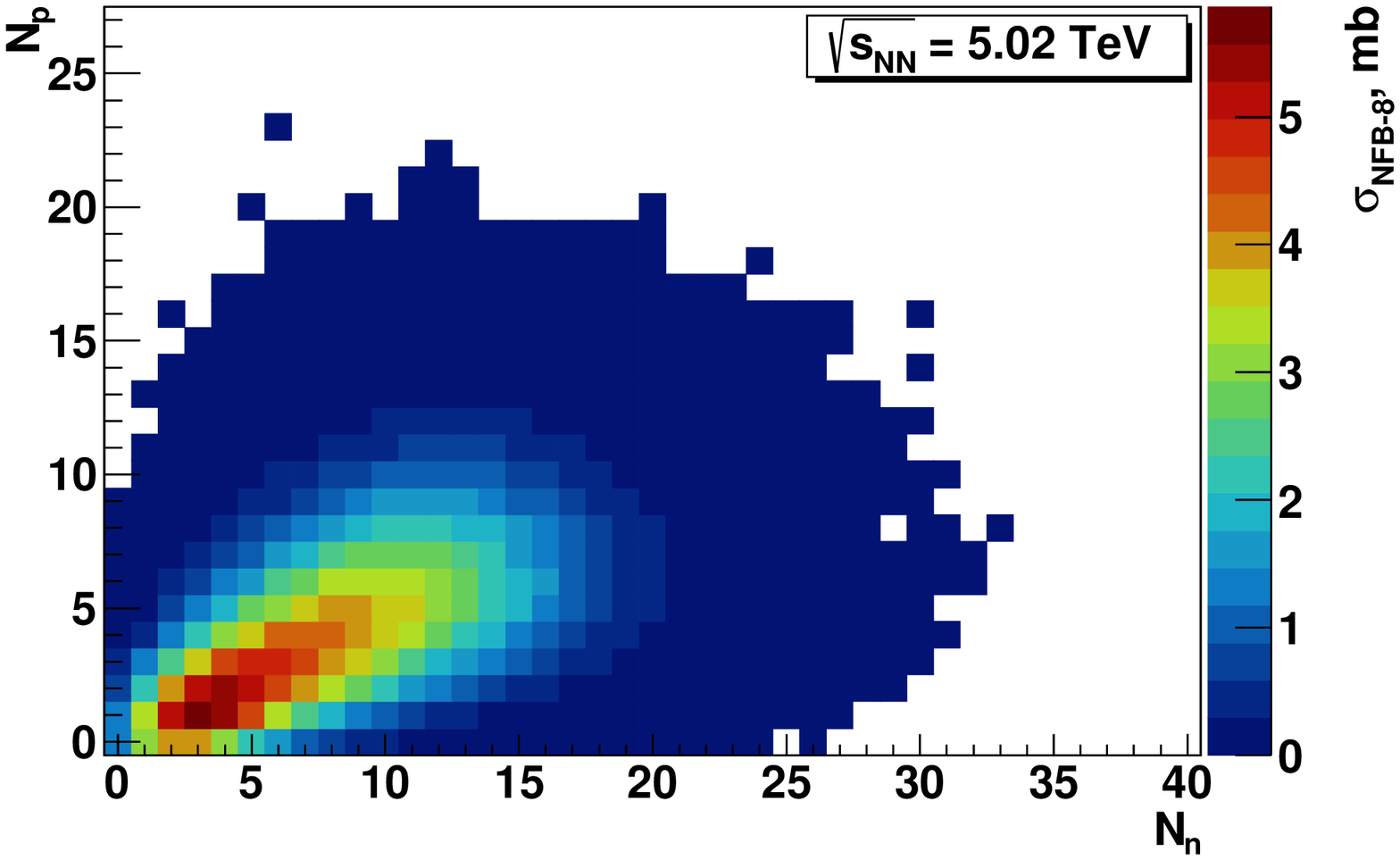}}
\end{subfigure}
\caption{Cross sections $\sigma(N_n, N_p)$ of emission of given numbers of spectator neutrons $N_n$ and protons $N_p$ in $^{208}$Pb--$^{208}$Pb collisions for 0--5\% centrality at $\sqrt{s_{\mathrm{NN}}}=17.21$~GeV (top) and $\sqrt{s_{\mathrm{NN}}}=5.02$~TeV (bottom) calculated with NFB-8 nuclear density profile in $^{208}$Pb.}
\label{fig:np_distr}
\end{figure}

The reduction in the numbers of $N_n$ and $N_p$ at the LHC energy is also seen in the partial cross sections $\sigma(N_n, N_p)$ of emission of given numbers of neutrons and protons calculated with AAMCC-MST with NFB-8 profile for collisions at 0--5\% centrality. These cross sections are presented in Fig.~\ref{fig:np_distr} for two collision energies. As follows from calculations, the maximum of $\sigma(N_n, N_p)\sim 3$~mb at the SPS is predicted for the  emission of 11--17 neutrons accompanied by 6--9 protons. However, the maximum of $\sigma(N_n, N_p)\sim 6$~mb at the LHC is obtained for the emission of only 2--4 neutrons along with 1--2 protons. While in the simulated ultracentral $^{208}$Pb--$^{208}$Pb collisions about 25 spectator neutrons and 15 spectator protons are emitted with noticeable cross sections at the SPS, such high multiplicities of spectator nucleons are very improbable in collisions at the LHC, see Fig.~\ref{fig:np_distr}.

\begin{figure}[tbp!]
\begin{subfigure}[b]{1.0\linewidth}
\centerline{\includegraphics[width=1.0\linewidth]{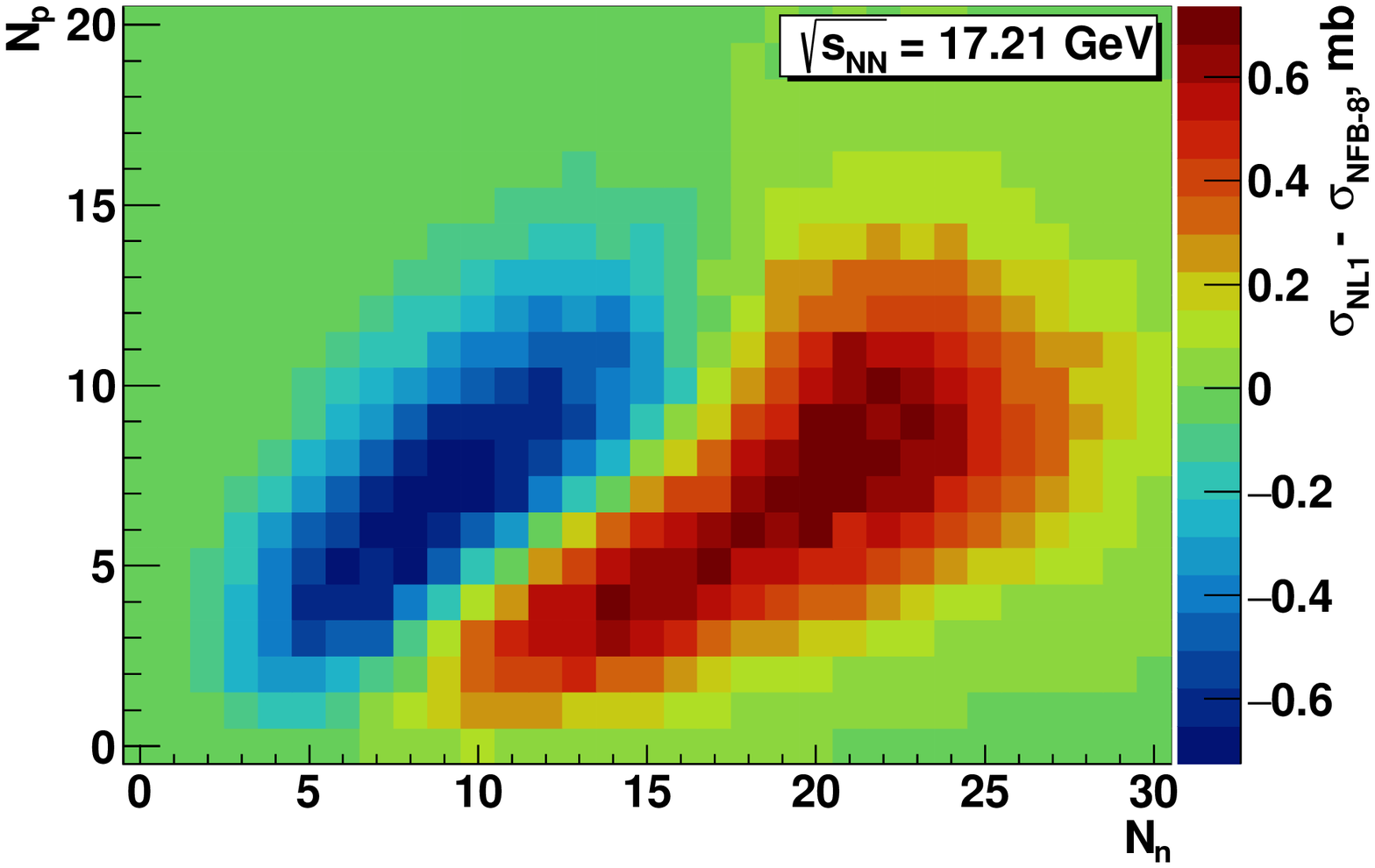}}
\end{subfigure}
\begin{subfigure}[b]{1.0\linewidth}
\centerline{\includegraphics[width=1.0\linewidth]{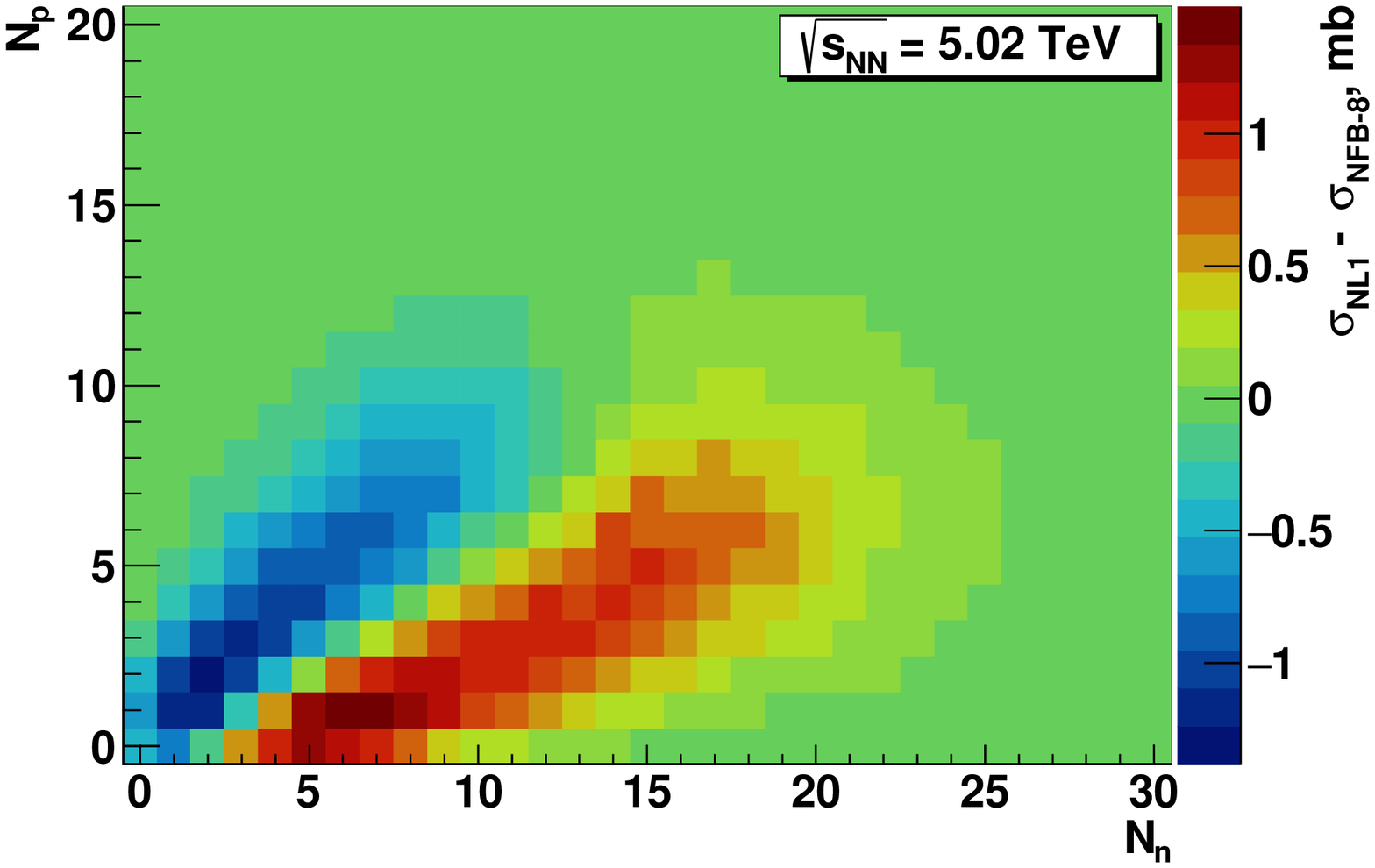}}
\end{subfigure}
\caption{Difference between the cross sections $\sigma(N_n, N_p)$ of emission of given numbers of spectator neutrons $N_n$ and protons $N_p$ calculated with NL1 and NFB-8 nuclear density profiles for $^{208}$Pb--$^{208}$Pb collisions for 0--5\% centrality at  $\sqrt{s_{\mathrm{NN}}}=17.21$~GeV (top) and $\sqrt{s_{\mathrm{NN}}}=5.02$~TeV (bottom).}
\label{fig:np_distr_diff}
\end{figure}

The main goal of the present work consists in evaluating the sensitivity of the parameters of NS to the yields of spectrator neutrons and protons in ultracentral $^{208}$Pb--$^{208}$Pb collisions. With this in mind the cross sections $\sigma(N_n, N_p)$ were calculated not only with NFB-8 nuclear density profiles representing a thin NS as in Fig.~\ref{fig:np_distr}, but also with NL1 corresponding to thick NS. The changes in $\sigma(N_n, N_p)$ caused by the replacement of NFB-8 by NL1 in calculations can be seen from Fig.~\ref{fig:np_distr_diff} where the absolute difference $\sigma_{\rm NL1}-\sigma_{\rm NFB-8}$ between two calculations is presented as function of $N_n$ and $N_p$. Also in this case, the results are given at  $\sqrt{s_{\mathrm{NN}}}=17.21$~GeV and 5.02~TeV to understand the dependence on collision energy.

As seen from Fig.~\ref{fig:np_distr_diff}, additional surface neutrons in NL1 lead to larger $\sigma(N_n, N_p)$ for $N_n$ between 10 and 27 and $N_p$ between 2 and 14 in $^{208}$Pb--$^{208}$Pb collisions at the SPS. Qualitatively, a similar increase of $\sigma(N_n, N_p)$ is seen also for collisions at the LHC, but for  smaller numbers of spectator nucleons: $N_n=4-18$ and $N_p=0-7$. More neutrons and less protons are obtained in general  in calculations with NL1 representing a thick NS  ($\Delta r_{np}=0.321$~fm) in comparison to NFB-8 option ($\Delta r_{np}=0.115$~fm). 

\begin{figure*}[htbp!]
\centerline{\includegraphics[width=1.0\linewidth]{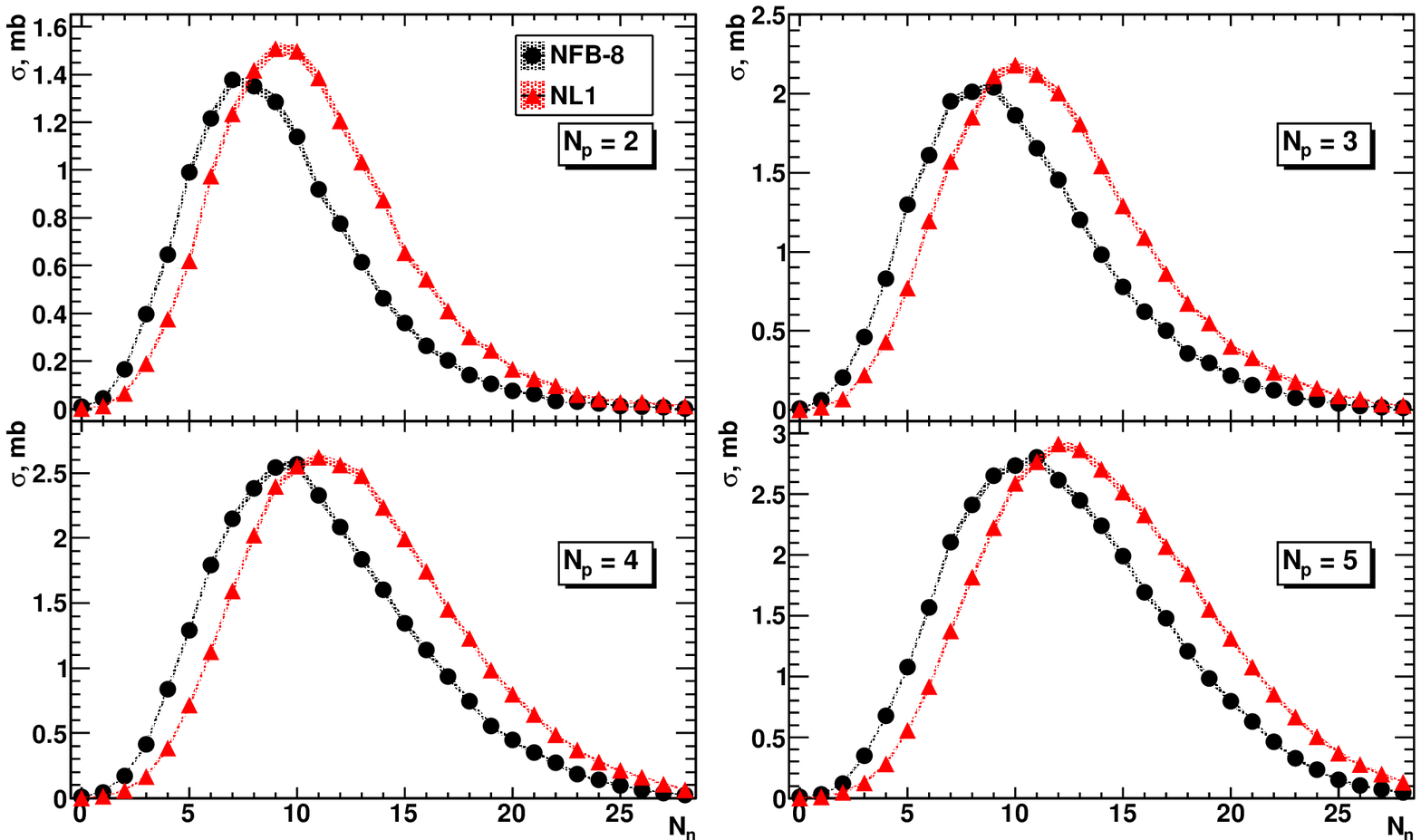}}
\caption{Cross sections $\sigma(N_n,N_p)$ of emission of given numbers of spectator neutrons $N_n$ in events with specific number of spectator protons $N_p=$2,~3,~4 or 5 in $^{208}$Pb--$^{208}$Pb collisions of 0--5\% centrality at $\sqrt{s_{\mathrm{NN}}}=17.21$~GeV calculated with NFB-8 and NL1 density profiles in $^{208}$Pb.}
\label{fig:n_p_fixed}
\end{figure*}
\begin{figure}[tb!]
\centerline{\includegraphics[width=1.0\linewidth]{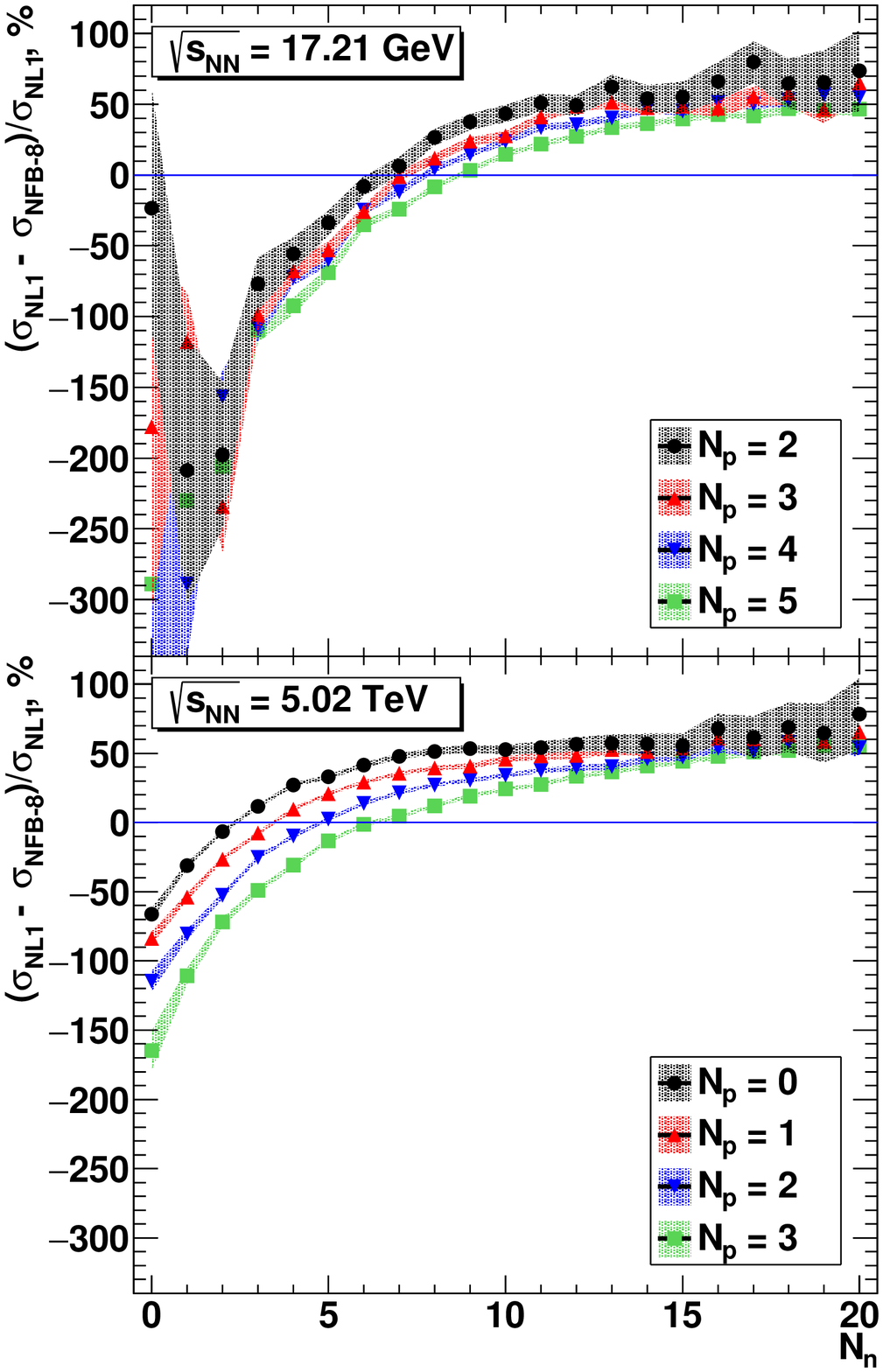}}
\caption{Relative difference between the cross sections $\sigma(N_n,N_p)$ of emission of a given  number of spectator neutrons $N_n$ accompanied by a specific number of spectator protons $N_p$ calculated with NL1 and NFB-8 profiles in $^{208}$Pb--$^{208}$Pb collisions for 0--5\% centrality. Top panel: at $\sqrt{s_{\mathrm{NN}}}=17.21$~GeV for $N_p=2,3,4,5$. Bottom panel: at  $\sqrt{s_{\mathrm{NN}}}=5.02$~TeV for $N_p=0,1,2,3$. Shadowed bands represent statistical uncertainties of calculations.}
\label{fig:n_p_fixed_rel}
\end{figure}

It is challenging to measure $\left\langle N_n \right\rangle$ and $\left\langle N_p \right\rangle$ in experiments because of expected wide ranges of $N_n$ and $N_p$ presented in  Fig.~\ref{fig:np_distr}. Events of low and high spectator nucleon multiplicity have to be detected. With a limited acceptance and efficiency of forward detectors sophisticated corrections for spectator nucleons lost in multinucleon events are necessary, as explained, in particular, in Ref.~\cite{Dmitrieva2018}. According to Table~\ref{tab:nav_en} only $\sim 15$\% differences in $\left\langle N_n \right\rangle$ are expected for thin and thick NS, and such a difference may be hidden in the uncertainties of detection of multinucleon events caused by efficiency limitations.  Therefore, the cross sections $\sigma(N_n, N_p)$ for relatively small $N_n$ and $N_p$ can be considered as more sensitive probes of the parameters of NS. In selecting suitable  intervals of $N_n$ and $N_p$ one can turn to 
Fig.~\ref{fig:np_distr_diff} to find neutron and proton multiplicities with the most prominent changes of $\sigma(N_n, N_p)$ caused by variations of the NS thickness in calculations. In particular, as seen from Fig.~\ref{fig:np_distr_diff}, it is not reasonable to plan measurements of the sum of $\sigma(N_n, N_p)$ over a wide range of $N_n=1-15$, but for specific numbers of $N_p=0,1,2,...5$. Indeed, the differences between $\sigma(N_n, N_p)$ calculated with NL1 and NFB-8 profiles have different signs, and thus their contributions to this sum compensate each other. This is seen from Fig.~\ref{fig:n_p_fixed} presenting $\sigma(N_n, N_p)$ calculated for  $^{208}$Pb--$^{208}$Pb collisions at 0--5\% centrality and $\sqrt{s_{\mathrm{NN}}}=17.21$~GeV in a wide range of $N_n$, but for specific $N_p=2,3,4$ or 5. The selection of $2 \le N_p \le 5$ is based on the difference between $\sigma(N_n, N_p)$ calculated with NL1 and NFB-8 shown in Fig.~\ref{fig:np_distr_diff}. Indeed, the largest absolute difference up to 0.6~mb is found for $2 \le N_p \le 5$ ultracentral $^{208}$Pb--$^{208}$Pb collisions at $\sqrt{s_{\mathrm{NN}}}=17.21$~GeV.  

As seen from Fig.~\ref{fig:n_p_fixed}, the maximum of $\sigma(N_n, N_p)$ calculated with a thicker NS corresponding to NL1 profile is shifted to larger $N_n$ compared to the calculations with NFB-8, but the sum of $\sigma(N_n, N_p)$ within the considered range of $N_n$ remains almost the same for calculations with NL1 and NFB-8. Therefore, the determination of $\sigma(N_n, N_p)$ for specific $N_n$ and $N_p$ can be proposed in future experiments, rather than measuring the sum of $\sigma(N_n, N_p)$ over $N_n$. 

The results for $\sigma(N_n, N_p)$ calculated with NL1 and NFB-8 for ultracentral $^{208}$Pb--$^{208}$Pb collisions at $\sqrt{s_{\mathrm{NN}}}=5.02$~TeV were found qualitatively similar to those shown in Fig.~\ref{fig:n_p_fixed} for $\sqrt{s_{\mathrm{NN}}}=17.21$~GeV, and they are not presented here. In comparison to lower collision energy, $\sigma(N_n, N_p)$ calculated at $\sqrt{s_{\mathrm{NN}}}=5.02$~TeV demonstrate a larger difference between the results obtained with NL1 and NFB-8 (up to 1.6~mb). However, this maximum value is found at lower $N_n$ and $N_p=0,1,2,3$ because of the above-described depletion of the spectator matter in $^{208}$Pb--$^{208}$Pb collisions caused by the increase of NN-cross section with collision energy.

The degree of sensitivity of $\sigma(N_n,N_p)$ to variations of the parameters of NS can be demonstrated by calculating the relative difference between $\sigma(N_n,N_p)$ obtained  with two different nuclear density profiles. The relative variation $(\sigma_{\mathrm{NL1}}-\sigma_{\mathrm{NFB-8}})/\sigma_{\mathrm{NL1}}$ calculated for NL1 and NFB-8 profiles as function of $N_n$ is presented in  Fig.~\ref{fig:n_p_fixed_rel} for $^{208}$Pb--$^{208}$Pb collisions $\sqrt{s_{\mathrm{NN}}}=17.21$~GeV and 5.02~TeV. For the considered two collision energies specific numbers of spectator protons were taken as $N_p=2,3,4,5$ and $N_p=0,1,2,3$, respectively, as explained above.

As seen from Fig.~\ref{fig:n_p_fixed_rel}, the value of  $(\sigma_{\mathrm{NL1}}-\sigma_{\mathrm{NFB-8}})/\sigma_{\mathrm{NL1}}$ drops to -250\% and -150\% for low $N_n$, and reaches 50\% and 60\% for high $N_n$ at $\sqrt{s_{\mathrm{NN}}}=17.21$~GeV and 5.02~TeV, respectively. This relative difference is larger for $N_p=5$ and $N_p=3$ in comparison to $N_p=2$ and $N_p=0$, respectively, for the considered collision energies. One can conclude that while $\sigma(N_n,N_p)$ are small for $N_n\le 7$ and $N_p\le 5$, precise measurements of these cross sections sensitive to the presence of NS in $^{208}$Pb in ultracentral $^{208}$Pb--$^{208}$Pb collisions can help to validate the predictions of nuclear structure models~\cite{Centelles2010}. Alternatively, larger  $\sigma(N_n,N_p)$ for $N_n>10$ and $N_p\le 5$ can be proposed for measurements at both collision energies for the same purpose. 

\subsection{Calculations with Pbpnrw and PREX profiles}
\label{Subsection_3_3}

\begin{figure}[tb!]
\centerline{\includegraphics[width=1.0\linewidth]{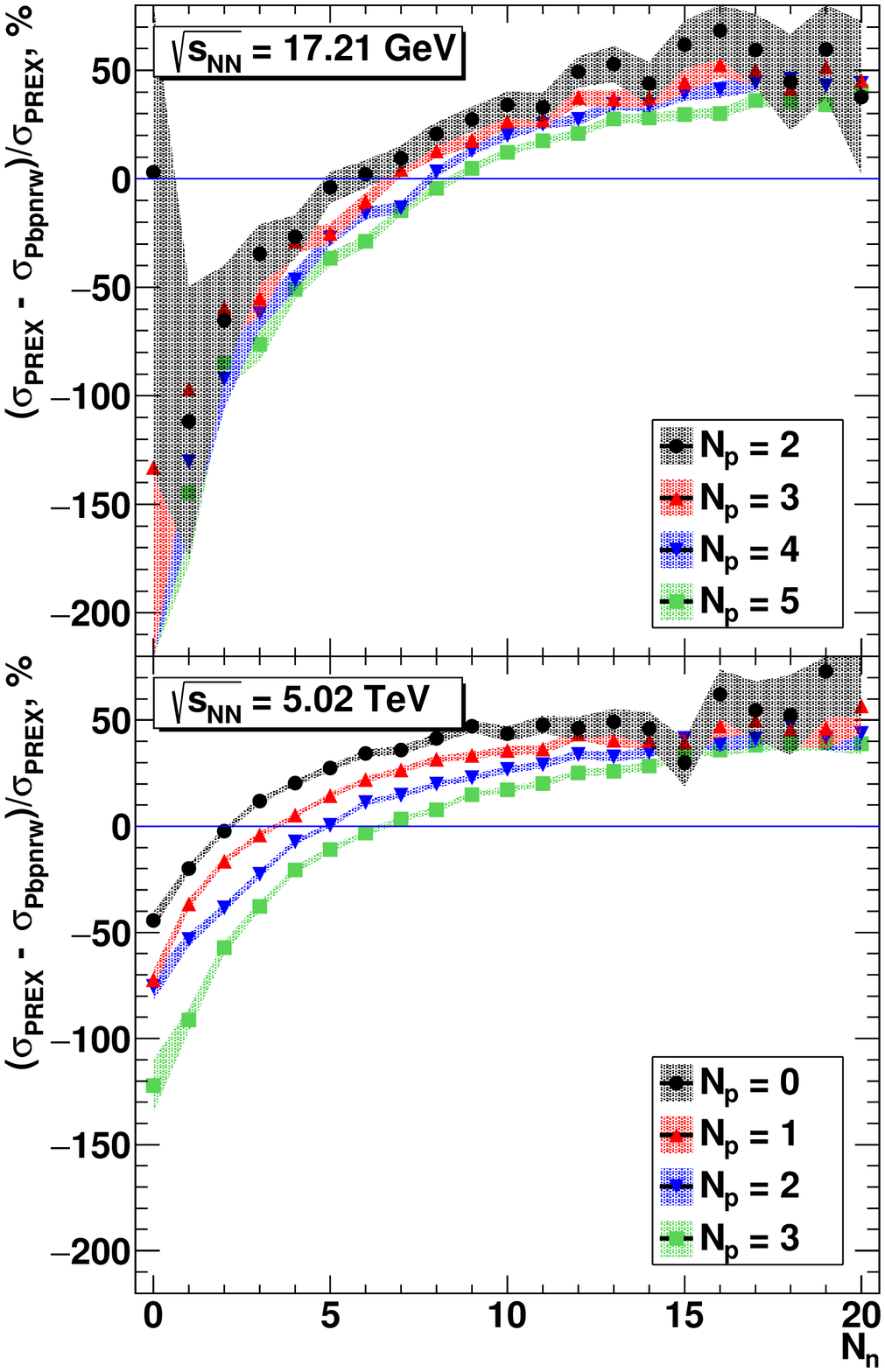}}
\caption{ Same as in Fig.~\ref{fig:n_p_fixed_rel}, but calculated with PREX and Pbpnrw nuclear density profiles.}
\label{fig:n_p_fixed_rel2}
\end{figure}

Calculations with AAMCC-MST also demonstrate that precise measurements of $\sigma(N_n,N_p)$ can help to confirm or refute the results of Crystal Ball at MAMI/A2 collaboration~\cite{Tarbert2014} and PREX~\cite{Adhikari2021} collaboration. For a quantitative assessment of the sensitivity of $\sigma(N_n,N_p)$ to the choice of nuclear density profiles, the ratio  $(\sigma_{\mathrm{PREX}}-\sigma_{\mathrm{Pbpnrw}})/\sigma_{\mathrm{PREX}}$ was calculated with results presented in Fig.~\ref{fig:n_p_fixed_rel2}.  

With respect to NS thickness, there is a similarity between AAMCC-MST results obtained with NFB-8 and Pbpnrw as profiles with thin NS ($\Delta r_{np}<0.15$~fm) on one side, and with thick NS  ($\Delta r_{np}>0.28$~fm) with NL1 and PREX profiles on the other side. This explains the similarity of results presented in Figs.~\ref{fig:n_p_fixed_rel} and \ref{fig:n_p_fixed_rel2}. However, the magnitude of variations caused by the replacement of Pbpnrw by PREX and reflected by  $(\sigma_{\mathrm{PREX}}-\sigma_{\mathrm{Pbpnrw}})/\sigma_{\mathrm{PREX}}$ is found somehow smaller in comparison to $(\sigma_{\mathrm{NL1}}-\sigma_{\mathrm{NFB-8}})/\sigma_{\mathrm{NL1}}$. It is typically at the level of -150\% and -80\% for small $N_n$ and up to 30\% and 50\% for high $N_n$ at $\sqrt{s_{\mathrm{NN}}}=17.21$~GeV and 5.02~TeV, respectively. Such a reduction of variations of  $\sigma(N_n, N_p)$ is expected because of a smaller difference in $\Delta r_{np}$ between PREX and Pbpnrw in comparison to the difference in $\Delta r_{np}$ between NL1 and NFB-8. Nevertheless, one can conclude that also in the former case precise measurements of $\sigma(N_n, N_p)$ for $N_n\le 7$ and $N_p\le 5$ sensitive to the presence of NS in $^{208}$Pb in ultracentral $^{208}$Pb--$^{208}$Pb collisions can help to validate the results of the  experiments~\cite{Tarbert2014,Adhikari2021}. Larger  $\sigma(N_n,N_p)$ calculated for $N_n>10$ and $N_p\le 5$ can be also proposed for measurements at both collision energies for the same purpose.

\subsection{Calculations with PREX1 and PREX2 profiles}
\label{Subsection_3_4}

As shown in Sections~\ref{Subsection_3_2} and \ref{Subsection_3_3}, the changes in the thickness $\Delta r_{np}$ of NS in calculations with AAMCC-MST lead to a significant change in $\sigma(N_n,N_p)$  for certain $N_n$ and $N_p$. In calculations with a thicker NS, the periphery of $^{208}$Pb contains more neutrons. As a result, $\sigma(N_n,N_p)$ in ultracentral collisions becomes larger for higher numbers $N_n$ of spectator neutrons.
\begin{figure}[tbp!]
\begin{subfigure}[b]{1.0\linewidth}
\centerline{\includegraphics[width=1.0\linewidth]{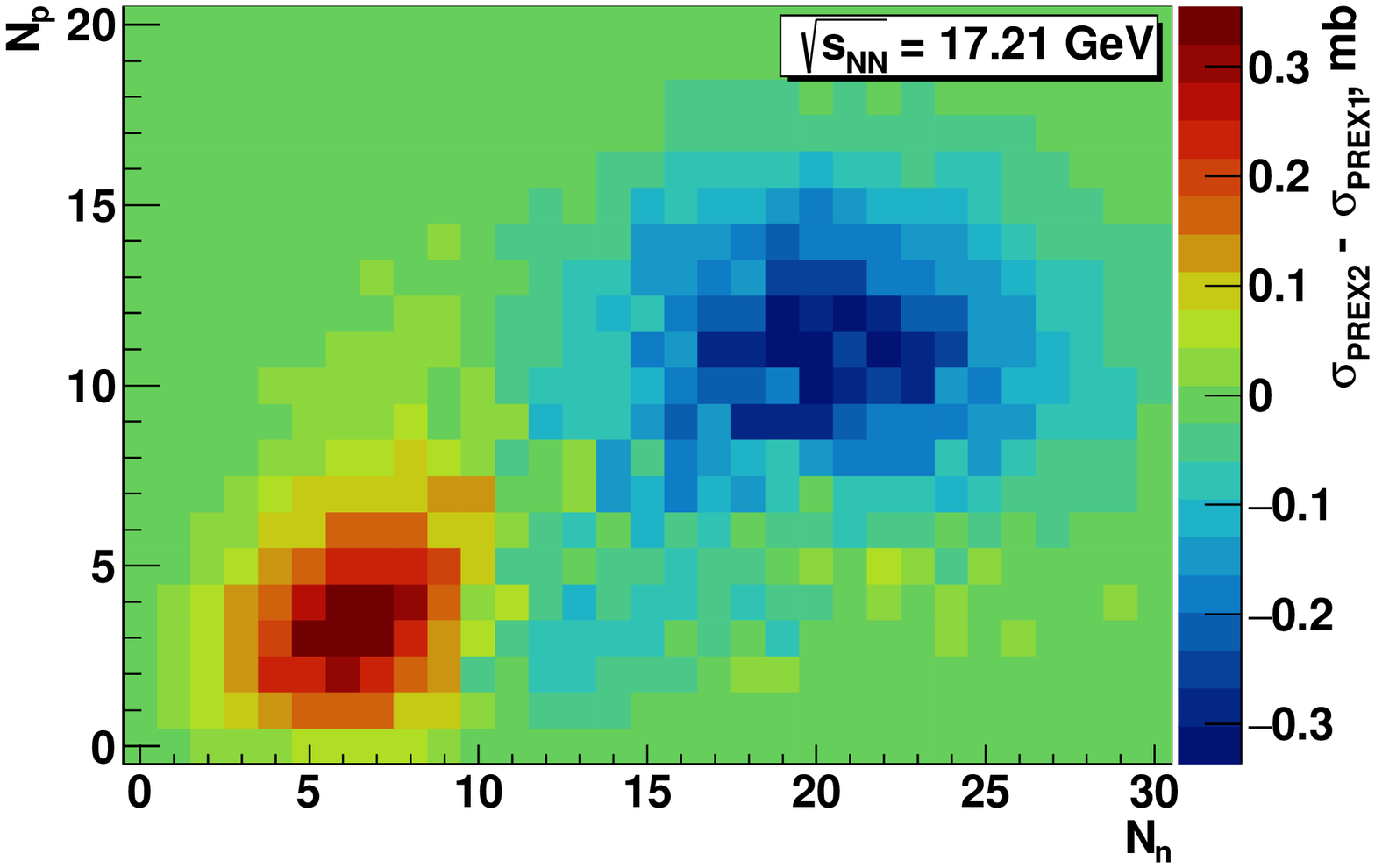}}
\end{subfigure}
\begin{subfigure}[b]{1.0\linewidth}
\centerline{\includegraphics[width=1.0\linewidth]{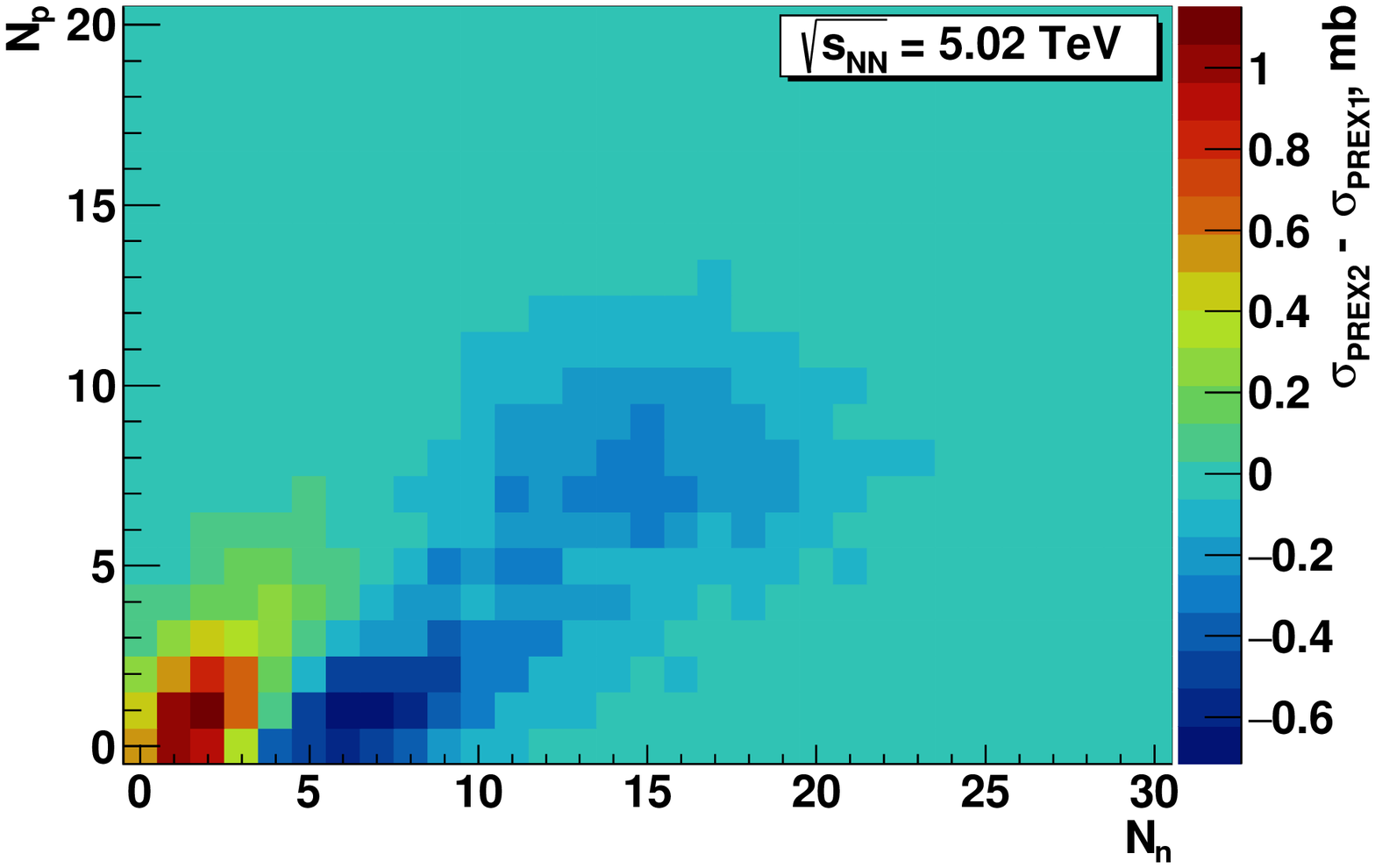}}
\end{subfigure}
\caption{Difference between the cross sections $\sigma(N_n, N_p)$ of emission of given numbers of spectator neutrons $N_n$ and protons $N_p$ calculated with PREX2 and PREX1 nuclear density profiles for $^{208}$Pb--$^{208}$Pb collisions for 0--5\% centrality at $\sqrt{s_{\mathrm{NN}}}=17.21$~GeV (top) and $\sqrt{s_{\mathrm{NN}}}=5.02$~TeV (bottom).}
\label{fig:np_distr_diff2}
\end{figure}

It is interesting to investigate whether the proposed method to constrain $\Delta r_{np}$ by measuring $\sigma(N_n, N_p)$ in ultracentral nucleus-nucleus collisions can also help in disentangling other fine details of nuclear periphery, like the relations between $R_n$, $R_p$, $a_n$ and $a_p$. With this purpose the difference in $\sigma(N_n, N_p)$  calculated with PREX2 ($R_n>R_p$, $a_n>a_p$) and PREX1 ($R_n=R_p$, $a_n>a_p$) profiles was calculated and presented in Fig.~\ref{fig:np_distr_diff2}. Both profiles are characterized by $\Delta r_{np}=0.283$~fm. However, as described above in Section~\ref{GlauberMC}, PREX1 profile represents the case of a neutron halo in $^{208}$Pb with a sufficiently large diffuseness parameter $a_n$ to compensate the equality $R_n=R_p$ and still provide the same $\Delta r_{np}=0.283$~fm as PREX2 profile with its $R_n>R_p$, see Table~\ref{tab:ns_par}.

\begin{figure}[tb!]
\centerline{\includegraphics[width=1.0\linewidth]{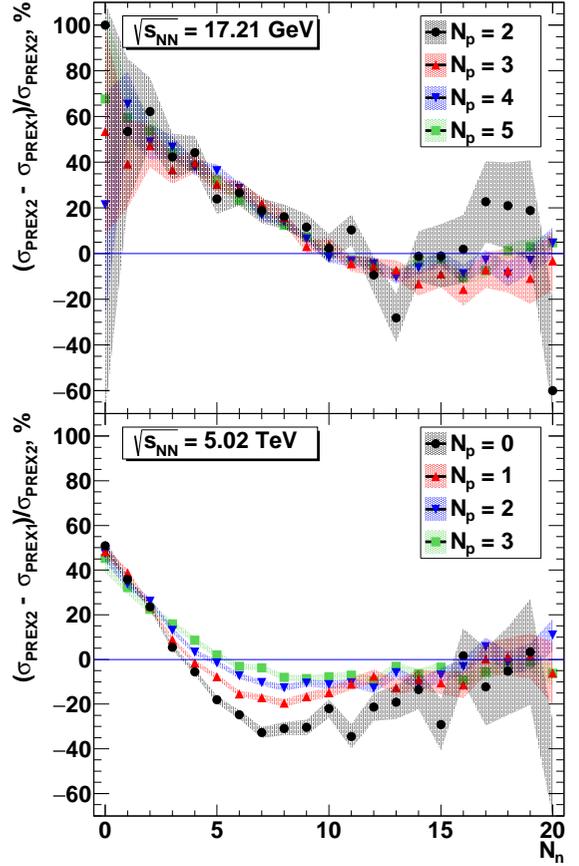}}
\caption{Same as in Fig.\ref{fig:n_p_fixed_rel}, but calculated with PREX2 and PREX1 nuclear density profiles.}
\label{fig:n_p_fixed_rel3}
\end{figure}

As seen from Fig.~\ref{fig:np_distr_diff2}, the difference between $\sigma(N_n, N_p)$ calculated with PREX2 and PREX1 profiles is smaller in comparison to the difference in cross sections calculated with NL1 and NFB-8 profiles presented in Fig.~\ref{fig:np_distr_diff}. This means that very precise measurements and calculations are necessary to understand whether theory and experiment are consistent with $R_n=R_p$ or rather with $R_n>R_p$. Nevertheless, one can conclude that the most sensitive to the presence or absence of NH cross sections $\sigma(N_n, N_p)$ are predicted basically in the intervals $N_n=1-10$ and $N_p=1-5$. The evolution of $\sigma(N_n, N_p)$ with collision energy, which is explained in Section~\ref{sect_2} by the depletion of spectator matter in collisions at the LHC, is seen also for the calculations with PREX1 and PREX2 profiles, see Fig.~\ref{fig:np_distr_diff2}. While noticeable differences in $\sigma(N_n, N_p)$ are confined within  $3<N_n<10$ and $N_p=2,3,4,5$ at $\sqrt{s_{\mathrm{NN}}}=17.21$~GeV, at high collision energy $\sqrt{s_{\mathrm{NN}}}=5.02$~TeV the differences are confined within $0<N_n<4$ and $N_p=0,1,2,3$.

Finally, the ratio  $(\sigma_{\mathrm{PREX2}}-\sigma_{\mathrm{PREX1}})/\sigma_{\mathrm{PREX2}}$ presented in Fig.~\ref{fig:n_p_fixed_rel3} can be considered at two collision energies. As seen, this ratio is up to 60\% at $\sqrt{s_{\mathrm{NN}}}=17.21$~GeV  and up to 50\% at $\sqrt{s_{\mathrm{NN}}}=5.02$~TeV. On one hand, large statistical uncertainties of calculations make it difficult to draw definite conclusions on the difference in the cross sections calculated with PREX2 and PREX1 profiles for $N_n>10$ at the LHC. On the other hand, in comparison to events with high multiplicity of spectator nucleons, events with $N_n\leq 3$ and $N_p\leq 3$ are preferable for detecting by ALICE Zero Degree Calorimeters because the corrections for ZDC efficiency are more simple for low nucleon multiplicities~\cite{Dmitrieva2018,Dmitrieva2021}

\subsection{Sensitivity to other parameters in calculations}\label{Subsection_3_5}

As shown above, the relative differences between the cross sections $\sigma(N_n, N_p)$ of emission of certain numbers of spectator nucleons in ultracentral $^{208}$Pb--$^{208}$Pb collisions calculated with different NS parameters are up to 250\% for certain $N_n$ and $N_p$. This demonstrates the sensitivity of calculated $\sigma(N_n, N_p)$ to the parameters of NS in $^{208}$Pb.  However, the NS parameters can be reliably constrained by the comparison of calculated and measured $\sigma(N_n, N_p)$ only if the changes of $\sigma(N_n, N_p)$ due to variations of other  parameters of the AAMCC-MST model are essentially smaller. 

The systematic uncertainties of the numbers of binary collisions $N_{coll}$ and participants $N_{part}$ calculated with Glauber MC model were estimated below  5\%, see Ref.~\cite{Loizides2018} for details. The combined uncertainties of $N_{coll}$ and $N_{part}$ due to the choice of the minimal internucleon separation, the algorithm of placing nucleon centers in nuclei (recentering) and $\sigma^\mathrm{NN}_{inel}$ were thoroughly evaluated. Therefore, in the present work only the sensitivity of $\sigma(N_n,N_p)$ to the main AAMCC-MST parameters and assumptions with a direct influence on the emission of forward spectator nucleons in ultracentral collisions is investigated. 

The results presented above in Sec.~\ref{Subsection_3_1}--Sec.~\ref{Subsection_3_4} were obtained by simulating the nucleon-nucleon collisions assuming the hard-sphere shape for nucleons, Eq.~(\ref{eq:p_b_default}). In Glauber MC model nucleon-nucleon collisions can be also simulated taking into account the fluctuations of the nucleon shape or partonic degrees of freedom~\cite{Loizides2016}, Eq.~(\ref{eq:p_b}). In particular, in Ref.~\cite{Loizides2018} $w=0.4$ was suggested leading to a Gaussian shape of this collision probability distribution. Therefore, it is necessary to evaluate the sensitivity of $\sigma(N_n, N_p)$ to the replacement of the hard-sphere option ($w=0$) by the fluctuating shape option ($w=0.4$) for NN collisions. For this purpose $^{208}$Pb--$^{208}$Pb collisions of 0--5\% centrality at $\sqrt{s_{\mathrm{NN}}}=5.02$~TeV were modelled with PREX nuclear density profile employing the both options. 
\begin{figure}[ht!]
\centerline{\includegraphics[width=1.0\linewidth]{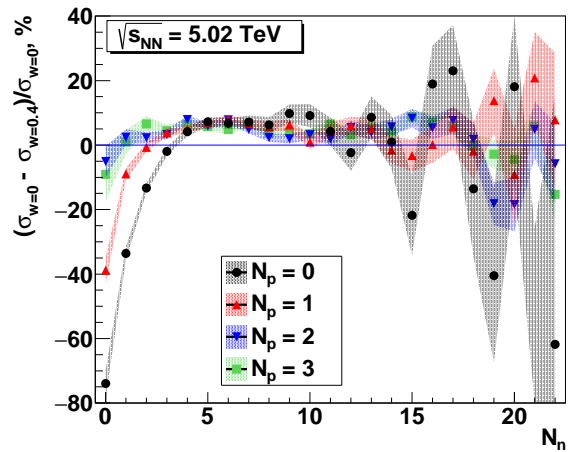}}
\caption{Relative difference between the cross sections $\sigma(N_n, N_p)$ of emission of given numbers of spectator neutrons $N_n$ and protons $N_p$ calculated with $w = 0$ and $w = 0.4$ for $^{208}$Pb--$^{208}$Pb collisions of 0--5\% centrality at $\sqrt{s_{\mathrm{NN}}}=5.02$~TeV. PREX nuclear density profile was used.}
\label{fig:w_inf}
\end{figure}

\begin{figure*}[ht!]
\centerline{\includegraphics[width=1.0\linewidth]{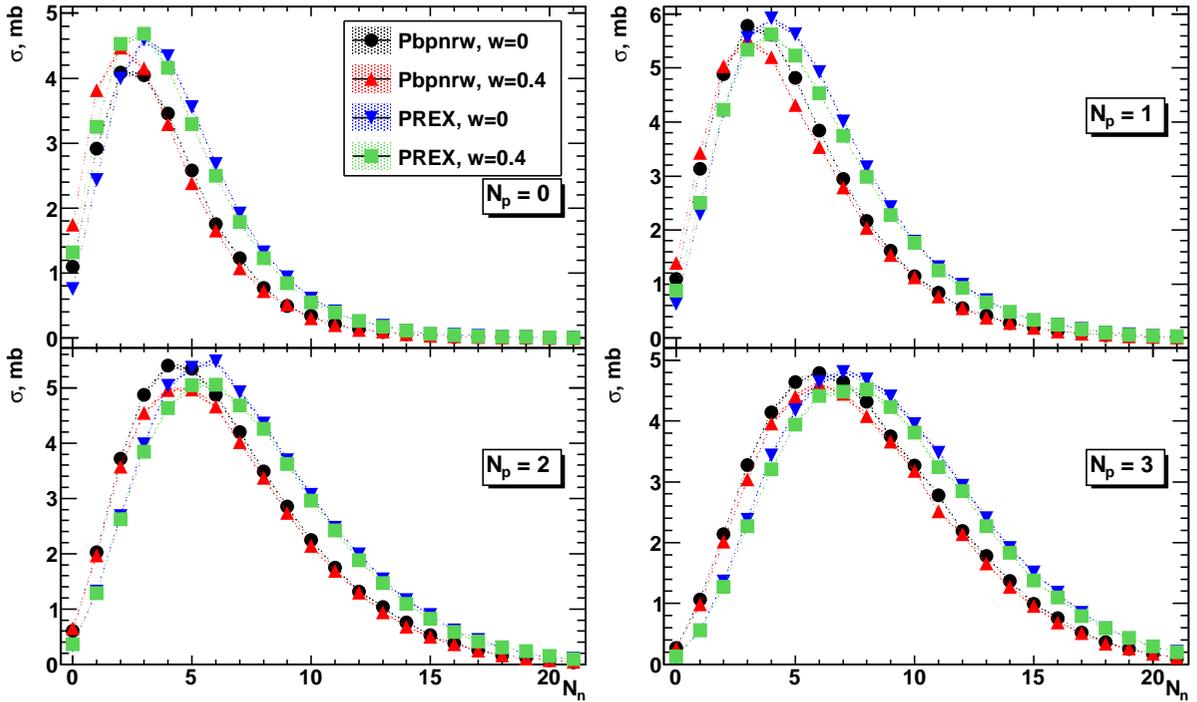}}
\caption{Cross sections $\sigma(N_n,N_p)$ of emission of given numbers of spectator neutrons $N_n$ in events with specific number of spectator protons $N_p=$ 0, 1, 2 or 3 in ultracentral $^{208}$Pb--$^{208}$Pb collisions for 0--5\% centrality at $\sqrt{s_{\mathrm{NN}}}=5.02$~TeV calculated with Pbpnrw and PREX density profiles in $^{208}$Pb with nominal hard-sphere $w = 0$ and  Gaussian $w = 0.4$ options in calculating NN-collision probability.}
\label{fig:w_absolute_cs}
\end{figure*}
The relative difference between the cross sections obtained with $w = 0$ and $w = 0.4$ is presented in Fig.~\ref{fig:w_inf}. 
As seen, the transition from $w = 0$ to $w = 0.4$ in Eq.~(\ref{eq:p_b}) leads to a noticeable increase in $\sigma(N_n, N_p)$ for $N_n =$ 0,~1 and $N_p =$ 0,~1 up to 80\%, and it is comparable to the difference between the respective cross sections calculated with PREX and Pbpnrw nuclear density profiles presented in  Fig.~\ref{fig:n_p_fixed_rel2}. Nevertheless, much smaller relative changes of $\sim 10$\% are seen for $\sigma(N_n, N_p)$ for $N_p =$ 2,~3 and intermediate neutron multiplicities $2<N_n<15$. Such changes are smaller than the relative difference ($\pm 50$\%) between the cross sections obtained with PREX and Pbpnrw nuclear density profiles for the same multiplicities of spectator neutrons and protons, see Fig.~\ref{fig:n_p_fixed_rel2}. Despite of the variations in modelling nucleon-nucleon collisions caused by replacing $w = 0$ with $w = 0.4$, the absolute values of $\sigma(N_n, N_p)$ calculated for $N_p =$0, 1, 2, 3 and $7<N_n<15$ with PREX and Pbpnrw profiles in ultracentral collisions still remain distinguishable, as shown in Fig.~\ref{fig:w_absolute_cs}. The conclusion that the details of the NN interaction model play a minor role in central collisions is in line with findings of Ref.~\cite{Alvioli2012}.

\begin{table}[ht!]
\centering
\caption{Average numbers of spectator neutrons $\left\langle N_n \right\rangle$, protons $\left\langle N_p \right\rangle$ and deuterons $\left\langle N_d \right\rangle$ in ultracentral ($b = 2$ fm) $^{208}$Pb--$^{208}$Pb collisions at $\sqrt{s_{\mathrm{NN}}} = 17.21$~GeV calculated with AAMCC-MST with nominal hard-sphere $w = 0$ and  Gaussian $w = 0.4$ options in calculating NN collision probability and without and with accounting for the nucleon-nucleon correlations. No neutron skin was assumed in all calculations, $\Delta r_{np}=0$. NA49 data~\cite{Appelshauser1998} are given for comparison.}
\begin{tabular}{|c|ccc|}
\hline
  & $\left\langle N_n \right\rangle$ & $\left\langle N_p  \right\rangle$ & $\left\langle  N_d \right\rangle$ \\
 \hline
hard-sphere, uncorrelated  & $12.13$  &  $7.48$  &  $0.82$  \\
hard-sphere,  correlated  & $11.87$  &  $7.41$  &  $0.67$  \\
\hline
 Gaussian,  uncorrelated  & $12.99$  &  $8.03$  &  $0.91$  \\
 Gaussian,  correlated  & $12.72$  &  $7.91$  &  $0.75$  \\
\hline
NA49  & $9.0 \pm 1.8$  &  $7.0 \pm 1.4$  &  0.5 \\
\hline
\end{tabular}
\label{tab:NN_w}
\end{table}

In Ref.~\cite{Alvioli2011} the production of spectator nucleons in collisions of ultrarelativistic nuclei was modelled taking into account short-range nucleon-nucleon correlations in nuclei. The authors of Refs.~\cite{Alvioli2009,Alvioli2011} made publicly available the sets of configurations of nucleons in several nuclei, including $^{208}$Pb, generated with accounting for the nucleon-nucleon correlations. In dedicated AAMCC-MST runs these configurations were introduced as initial positions of nucleons in colliding nuclei to model $^{208}$Pb--$^{208}$Pb collisions for 0--5\% centrality at $\sqrt{s_{\mathrm{NN}}}=5.02$~TeV. Then, the results of calculations with accounting for the NN correlations can be compared with results obtained with uncorrelated nucleons. In the available configurations of nucleons~\cite{Alvioli2009,Alvioli2011} the effects of NS were apparently neglected. For consistency, in respective AAMCC-MST calculations disregarding the NN correlation the values of the half-density radii and diffuseness parameters were set equal for neutrons and protons: $R_{n}=R_{p}=6.624$~fm and $a_{n}=a_{p}=0.549$~fm. This resulted in $\Delta r_{np}=0$ in calculations with and without NN correlations. In the latter case also neglecting NS, nucleons were placed according to the standard procedure adopted in the Glauber MC model~\cite{Loizides2018}. 

\begin{figure}[ht!]
\centerline{\includegraphics[width=1.0\linewidth]{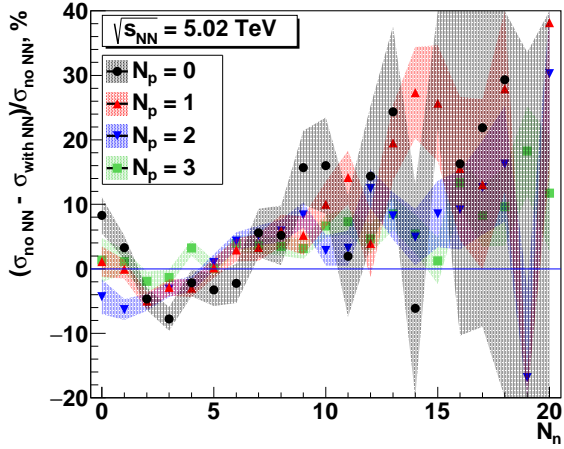}}
\caption{Relative difference between the cross sections $\sigma(N_n, N_p)$ of emission of given numbers of spectator neutrons $N_n$ and protons $N_p$ calculated without and with accounting for  nucleon-nucleon correlations for $^{208}$Pb--$^{208}$Pb collisions of 0--5\% centrality at $\sqrt{s_{\mathrm{NN}}}=5.02$~TeV. No neutron skin was assumed in all calculations, $\Delta r_{np}=0$.}
\label{fig:NN_inf}
\end{figure}

With the results presented in Table~\ref{tab:NN_w} and Fig.~\ref{fig:NN_inf} the impact of accounting for the NN correlations in nuclei on the yields of spectator neutrons, protons and deuterons can be estimated for ultracentral $^{208}$Pb--$^{208}$Pb collisions at $\sqrt{s_{\mathrm{NN}}}=17.2$~GeV and also at $5.02$~TeV. As seen from Table~\ref{tab:NN_w}, the average numbers of spectator neutrons $\left\langle N_n \right\rangle$ and protons $\left\langle N_p \right\rangle$ are changed by less than 1--2\% when these numbers are calculated with accounting for the NN correlations. Such small variations are also estimated in calculations  with accounting for the fluctuating shape of nucleons, with $w=0.4$ in Eq.~(\ref{eq:p_b}). The changes in the smaller average numbers of deuterons $\left\langle N_d \right\rangle$ are found larger, up to 18\%. In general, with the exception of the production of deuterons, the agreement with NA49 data~\cite{Appelshauser1998} is not improved by introducing the NN correlations.   

The relative difference between $\sigma(N_n, N_p)$ calculated without and with the nucleon-nucleon correlations for $^{208}$Pb--$^{208}$Pb collisions of 0--5\% centrality at $\sqrt{s_{\mathrm{NN}}}=5.02$~TeV is presented in Fig.~\ref{fig:NN_inf}.

As seen from Fig.~\ref{fig:NN_inf}, the changes in $\sigma(N_n, N_p)$ calculated with accounting for nucleon-nucleon correlation are within $\pm 10$\%  for all considered $N_p$ and for $N_n < 10$. While in these calculations the presence of NS in $^{208}$Pb was neglected, a similar sensitivity of $\sigma(N_n, N_p)$ to the NN correlations ($\sim 10$\%) can be expected in calculations with more realistic nuclear density profiles with NS,  Table~\ref{tab:ns_par}.

In AAMCC-MST, the yields of spectator neutrons and protons depend also on the critical distance  $d$ used in the MST-clustering algorithm presented in  Sec.~\ref{MST}. The parameter $d_0= 2.7$~fm is used in Eq.~(\ref{eq:clustdist}) to calculate $d$, and it is typically larger than the average internucleon distance  ($\sim 2$~fm) in nuclei in the ground state. While the NA49 data~\cite{Appelshauser1998} are described with $d_0= 2.7$~fm in general, see Table~\ref{tab:na49}, a smaller value $d_0 = 2.4$~fm can be also considered to evaluate the sensitivity of $\sigma(N_n, N_p)$ to $d_0$. 

In Table~\ref{tab:w_d0} the average numbers of spectator neutrons $\left\langle N_n \right\rangle$ and protons $\left\langle N_p \right\rangle$ calculated for $^{208}$Pb--$^{208}$Pb collisions of 0--5\% centrality at $\sqrt{s_{\mathrm{NN}}} = 5.02$~TeV are presented. In this table the results are divided into two groups representing, respectively, calculations with Pbpnrw and PREX nuclear density profiles. Some variations of $\left\langle N_n \right\rangle$ up to 5\% are seen for each nuclear density profile depending on $w$ and $d_0$ used in calculations. However, the difference in $\left\langle N_n \right\rangle$, and partially in $\left\langle N_p \right\rangle$, obtained with Pbpnrw and PREX is preserved. This means that in ultracentral collisions $\left\langle N_n \right\rangle$ and $\left\langle N_p \right\rangle$ are defined mainly by the specific nuclear density profile used in modelling with AAMCC-MST rather than by other calculational parameters. 

\begin{table}[ht!]
\centering
\caption{Average numbers of spectator neutrons $\left\langle N_n \right\rangle$ and protons $\left\langle N_p \right\rangle$ in  $^{208}$Pb--$^{208}$Pb collisions of 0--5\% centrality at $\sqrt{s_{\mathrm{NN}}} = 5.02$~TeV calculated with AAMCC-MST for Pbpnrw and PREX nuclear density profiles with various parameters in modelling NN collisions (hard-sphere, $w=0$ and Gaussian, $w=0.4$) and in estimating critical distance in MST-clusering ($d_0=2.7$~fm and $d_0=2.4$~fm).}
\begin{tabular}{|l|cc|}
\hline
  & $\left\langle N_n \right\rangle$ & $\left\langle N_p  \right\rangle$  \\
 \hline
Pbpnrw, hard-sphere, $d_0=2.7$~fm  & 8.96 & 4.59 \\
Pbpnrw, Gaussian, $d_0=2.7$~fm  & 9.34  &  4.8  \\
Pbpnrw, hard-sphere, $d_0=2.4$~fm  & 9.39 & 4.98 \\
\hline
PREX, hard-sphere, $d_0=2.7$~fm & 9.7  &  4.34   \\
PREX, Gaussian, $d_0=2.7$~fm  & 10.19  &  4.61   \\
PREX, hard-sphere, $d_0=2.4$~fm  & 10.11  & 4.71  \\
\hline
\end{tabular}
\label{tab:w_d0}
\end{table}

\begin{figure}[ht!]
\centerline{\includegraphics[width=1.0\linewidth]{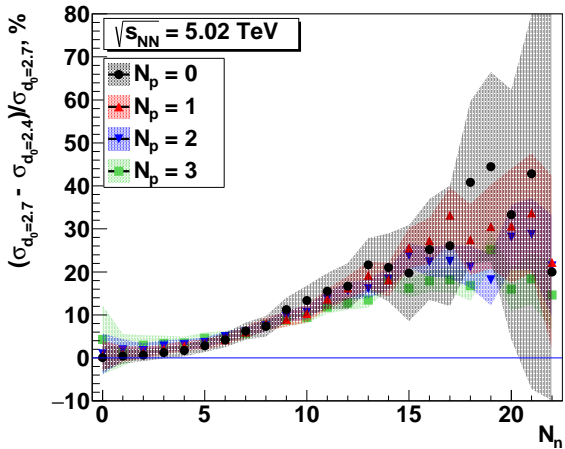}}
\caption{Relative difference between the cross sections $\sigma(N_n, N_p)$ of emission of given numbers of spectator neutrons $N_n$ and protons $N_p$ calculated with $d_0 = 2.7$ fm and $d_0 = 2.4$ fm for $^{208}$Pb--$^{208}$Pb collisions of 0--5\% centrality at $\sqrt{s_{\mathrm{NN}}}=5.02$~TeV. PREX density profile was used.}
\label{fig:d_inf}
\end{figure}

In Fig.~\ref{fig:d_inf} the relative difference between $\sigma(N_n, N_p)$ calculated for $^{208}$Pb--$^{208}$Pb collisions of 0--5\% centrality at $\sqrt{s_{\mathrm{NN}}}=5.02$~TeV  with the nominal  $d_0 = 2.7$~fm and reduced $d_0 = 2.4$~fm is presented. PREX density profile was used in this modelling. As seen from Fig.~\ref{fig:d_inf}, the variation of $\sigma(N_n, N_p)$ for $N_n < 10$ and $N_p < 4$ is less than 10\%. This means that the  cross sections calculated for the specified  multiplicities of spectator nucleons are relatively stable with respect to variations of $d_0$. 

\begin{figure*}[htb!]
\centerline{\includegraphics[width=1.0\linewidth]{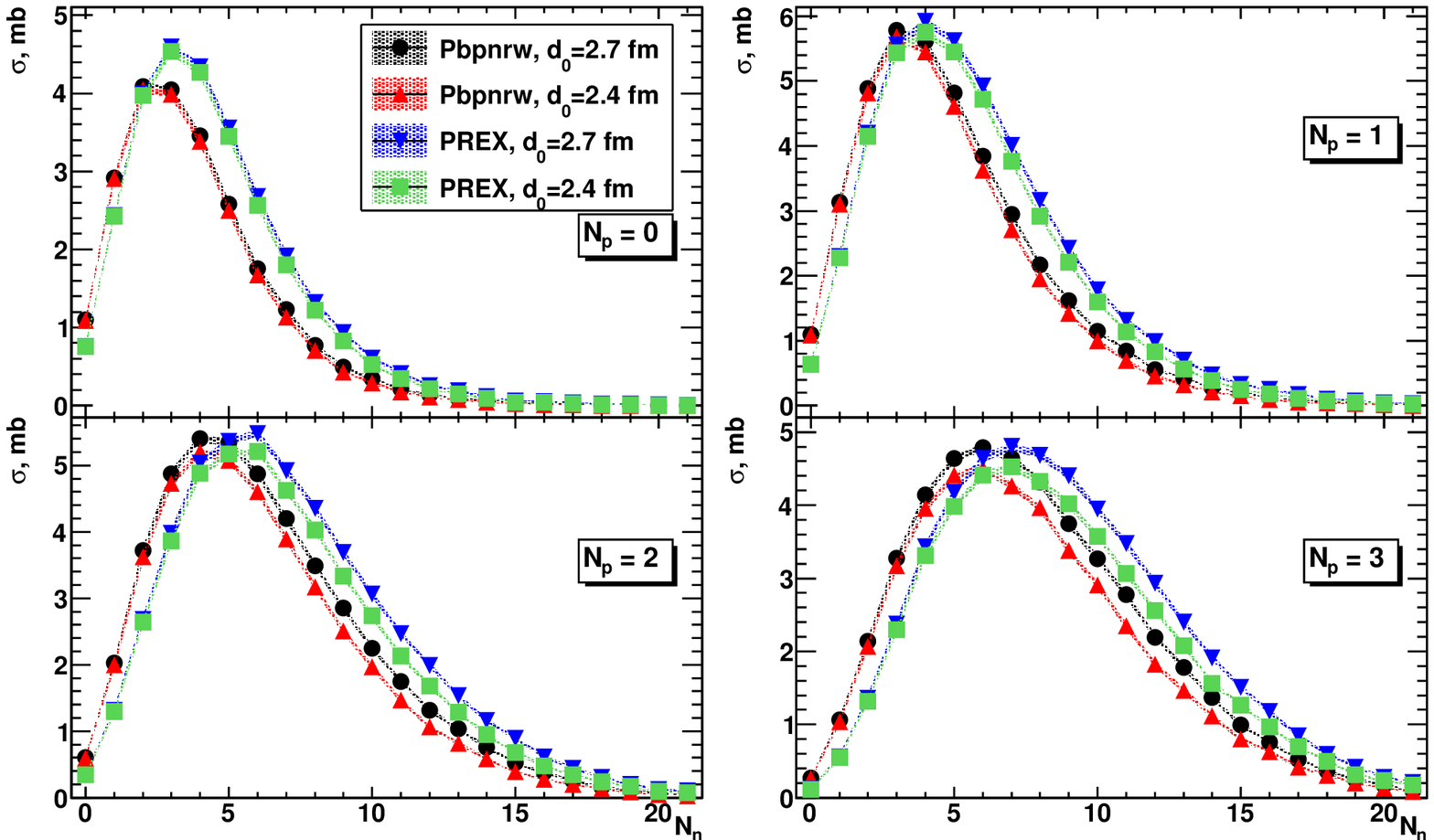}}
\caption{Cross sections $\sigma(N_n,N_p)$ of emission of given numbers of spectator neutrons $N_n$ in events with specific number of spectator protons, $N_p=$ 0, 1, 2 or 3, in $^{208}$Pb--$^{208}$Pb collisions of 0--5\% centrality at $\sqrt{s_{\mathrm{NN}}}=5.02$~TeV calculated with Pbpnrw and PREX density profiles in $^{208}$Pb with the nominal $d_0 = 2.7$~fm and reduced $d_0 = 2.4$~fm.}
\label{fig:d_absolute_cs}
\end{figure*}

In order to prove that the uncertainties of AAMCC-MST results due to the choice of $d_0$ still make possible to distinguish different NS parameterizations, calculations with the Pbpnrw profile were also performed with $d_0 = 2.7$~fm and $d_0 = 2.4$~fm. The results presented in Fig.~\ref{fig:d_absolute_cs} demonstrate that the values of $\sigma(N_n, N_p)$ remain well separated for $N_p=0$ and 1 for all $N_n$ despite of variations  of $d_0$. The cross sections with  $N_p=2$ and 3 remain distinguishable for $N_n < 5$. Therefore, the cross sections measured for the above-mentioned $N_n$ and $N_p$ can be safely used to constrain the parameters of NS by comparison with calculations.

\section{Conclusions}

As demonstrated by calculations with the AAMCC-MST model, the yields of spectator neutrons and protons in ultracentral (0--5\% centrality) $^{208}$Pb--$^{208}$Pb collisions at $\sqrt{s_{\mathrm{NN}}}=17.21$~GeV and  $5.02$~TeV are sensitive to the parameters of NS in $^{208}$Pb. Spectator matter in ultracentral $^{208}$Pb--$^{208}$Pb collisions is represented mostly by nucleons which are peeled away from the nuclear periphery representing NS. Depending on the thickness $\Delta r_{np}$ of NS, and on the relations between the half-density radii of the distributions of neutrons and protons ($R_n=R_p$ vs $R_n>R_p$), very modest changes of the average numbers of spectator neutrons were calculated. 

In contrast, the cross sections $\sigma(N_n, N_p)$ of emission of given numbers of spectator neutrons $N_n$ accompanied by specific number of spectator protons $N_p$ calculated with different parameterizations of NS diverge up to 250\% for certain numbers of $N_n$ and $N_p$. In particular, $\sigma(N_n, N_p)$ calculated for $N_n=1,2...7$ and $N_p=0,1,2,3$ in ultracentral $^{208}$Pb--$^{208}$Pb collisions at the LHC were found to be very sensitive to $\Delta r_{np}$ in $^{208}$Pb. Accurate measurements of $\sigma(N_n, N_p)$ may also help to understand whether theory and experiment are consistent with $R_n=R_p$ or rather with $R_n>R_p$. The cross sections $\sigma(N_n, N_p)$ of emission of low numbers of spectator neutrons and protons ($N_n\leq 7$ and $N_p\leq 3$) can be proposed for measurements in the ALICE experiment at the LHC providing that the yields of spectator nucleons will be thoroughly corrected for the acceptance and efficiency of the ALICE Zero Degree Calorimeters~\cite{Puddu2007,Dmitrieva2018}. 
As reported by ALICE, the multiplicities of forward spectator neutrons and protons can be measured individually in events of different centralities~\cite{CollaborationALICE2020}, while the collisions of 0--5\% centrality can be identified by different well-established methods~\cite{Abelev2013}. The cross sections of emission of 1--3 forward neutrons resulting from electromagnetic dissociation of $^{208}$Pb at the LHC were already measured~\cite{Abelev2012n}.  

In order to summarize the studies of sensitivities of $\sigma(N_n, N_p)$ to the parameters of calculations considered in this work, one can conclude that $\sigma(N_n, N_p)$ for $N_p=$0,~1 are sufficiently stable ($\pm 10$\%) for intermediate $2<N_n<10$. The cross sections $\sigma(N_n, N_p)$ for $N_p=$2,~3 can be also proposed to probe NS effects, but for $N_n<5$. Their variations are also sufficiently small ($\sim 10$\%) in comparison to the difference in $\sigma(N_n, N_p)$ ($\sim$50--250\%) calculated with 
different NS parameters.

In summary, one can expect that  $\sigma (N_n, N_p)$ measured in ultracentral $^{208}$Pb--$^{208}$Pb collisions will help to constrain the existing variety of the parameters of NS predicted by various nuclear structure models and estimated from experimental data for $^{208}$Pb. 

\section{Acknowledgments}
One of the authors (I.P.) is grateful to Dariusz Miskowiec and Chiara Oppedisano for the discussions which stimulated the investigation of the effects of neutron skin in ultrarelaticistic $^{208}$Pb--$^{208}$Pb collisions.

\bibliography{N_Kozyrev_Peeling_NS}


\begin{thebibliography}{56}
\ifx \bisbn   \undefined \def \bisbn  #1{ISBN #1}\fi
\ifx \binits  \undefined \def \binits#1{#1}\fi
\ifx \bauthor  \undefined \def \bauthor#1{#1}\fi
\ifx \batitle  \undefined \def \batitle#1{#1}\fi
\ifx \bjtitle  \undefined \def \bjtitle#1{#1}\fi
\ifx \bvolume  \undefined \def \bvolume#1{\textbf{#1}}\fi
\ifx \byear  \undefined \def \byear#1{#1}\fi
\ifx \bissue  \undefined \def \bissue#1{#1}\fi
\ifx \bfpage  \undefined \def \bfpage#1{#1}\fi
\ifx \blpage  \undefined \def \blpage #1{#1}\fi
\ifx \burl  \undefined \def \burl#1{\textsf{#1}}\fi
\ifx \doiurl  \undefined \def \doiurl#1{\url{https://doi.org/#1}}\fi
\ifx \betal  \undefined \def \betal{\textit{et al.}}\fi
\ifx \binstitute  \undefined \def \binstitute#1{#1}\fi
\ifx \binstitutionaled  \undefined \def \binstitutionaled#1{#1}\fi
\ifx \bctitle  \undefined \def \bctitle#1{#1}\fi
\ifx \beditor  \undefined \def \beditor#1{#1}\fi
\ifx \bpublisher  \undefined \def \bpublisher#1{#1}\fi
\ifx \bbtitle  \undefined \def \bbtitle#1{#1}\fi
\ifx \bedition  \undefined \def \bedition#1{#1}\fi
\ifx \bseriesno  \undefined \def \bseriesno#1{#1}\fi
\ifx \blocation  \undefined \def \blocation#1{#1}\fi
\ifx \bsertitle  \undefined \def \bsertitle#1{#1}\fi
\ifx \bsnm \undefined \def \bsnm#1{#1}\fi
\ifx \bsuffix \undefined \def \bsuffix#1{#1}\fi
\ifx \bparticle \undefined \def \bparticle#1{#1}\fi
\ifx \barticle \undefined \def \barticle#1{#1}\fi
\bibcommenthead
\ifx \bconfdate \undefined \def \bconfdate #1{#1}\fi
\ifx \botherref \undefined \def \botherref #1{#1}\fi
\ifx \url \undefined \def \url#1{\textsf{#1}}\fi
\ifx \bchapter \undefined \def \bchapter#1{#1}\fi
\ifx \bbook \undefined \def \bbook#1{#1}\fi
\ifx \bcomment \undefined \def \bcomment#1{#1}\fi
\ifx \oauthor \undefined \def \oauthor#1{#1}\fi
\ifx \citeauthoryear \undefined \def \citeauthoryear#1{#1}\fi
\ifx \endbibitem  \undefined \def \endbibitem {}\fi
\ifx \bconflocation  \undefined \def \bconflocation#1{#1}\fi
\ifx \arxivurl  \undefined \def \arxivurl#1{\textsf{#1}}\fi
\csname PreBibitemsHook\endcsname

\bibitem{Botvina1995}
\begin{barticle}
\bauthor{\bsnm{Botvina}, \binits{A.S.}},
\bauthor{\bsnm{Mishustin}, \binits{I.N.}},
\bauthor{\bsnm{Begemann-Blaich}, \binits{M.}}, \betal:
\batitle{Multifragmentation of spectators in relativistic heavy-ion reactions}.
\bjtitle{{Nucl. Phys. A}}
\bvolume{584},
\bfpage{737}--\blpage{756}
(\byear{1995}).
\doiurl{10.1016/0375-9474(94)00621-S}
\end{barticle}
\endbibitem

\bibitem{Appelshauser1998}
\begin{barticle}
\bauthor{\bsnm{Appelsh{\"a}user}, \binits{H.}},
\bauthor{\bsnm{B{\"a}chler}, \binits{J.}},
\bauthor{\bsnm{Bailey}, \binits{S.J.}}, \betal:
\batitle{{Spectator Nucleons in Pb+Pb Collisions at 158~A~GeV}}.
\bjtitle{Eur. Phys. J. A}
\bvolume{2},
\bfpage{383}--\blpage{390}
(\byear{1998}).
\doiurl{10.1007/s100500050135}
\end{barticle}
\endbibitem

\bibitem{Puddu2007}
\begin{barticle}
\bauthor{\bsnm{Puddu}, \binits{G.}},
\bauthor{\bsnm{Arnaldi}, \binits{R.}},
\bauthor{\bsnm{Chiavassa}, \binits{E.}}, \betal:
\batitle{{The zero degree calorimeters for the ALICE experiment}}.
\bjtitle{Nucl. Inst. Meth. A}
\bvolume{581},
\bfpage{397}--\blpage{401}
(\byear{2007}).
\doiurl{10.1016/j.nima.2007.08.013}
\end{barticle}
\endbibitem

\bibitem{Abelev2013}
\begin{barticle}
\bauthor{\bsnm{Abelev}, \binits{B.}},
\bauthor{\bsnm{Adam}, \binits{J.}},
\bauthor{\bsnm{Adamov{\'a}}, \binits{D.}}, \betal:
\batitle{{Centrality determination of Pb-Pb collisions at $\sqrt{s_{NN}}=2.76$
  TeV with ALICE}}.
\bjtitle{Phys. Rev. C}
\bvolume{88},
\bfpage{044909}
(\byear{2013}).
\doiurl{10.1103/PhysRevC.88.044909}
\end{barticle}
\endbibitem

\bibitem{Barrett1977}
\begin{bbook}
\bauthor{\bsnm{Barrett}, \binits{R.C.}},
\bauthor{\bsnm{Jackson}, \binits{D.F.}}:
\bbtitle{Nuclear Sizes and Structure},
p. \bfpage{566}.
\bpublisher{Clarendon Press},
\blocation{New York}
(\byear{1977})
\end{bbook}
\endbibitem

\bibitem{Tanihata1985}
\begin{barticle}
\bauthor{\bsnm{Tanihata}, \binits{I.}},
\bauthor{\bsnm{Hamagaki}, \binits{H.}},
\bauthor{\bsnm{Hashimoto}, \binits{O.}}, \betal:
\batitle{{Measurements of interaction cross sections and radii of He
  isotopes}}.
\bjtitle{Phys. Lett. B}
\bvolume{160},
\bfpage{380}--\blpage{384}
(\byear{1985}).
\doiurl{10.1016/0370-2693(85)90005-X}
\end{barticle}
\endbibitem

\bibitem{Hansen1995}
\begin{barticle}
\bauthor{\bsnm{Hansen}, \binits{P.G.}},
\bauthor{\bsnm{Jensen}, \binits{A.S.}},
\bauthor{\bsnm{Jonson}, \binits{B.}}:
\batitle{{Nuclear halos}}.
\bjtitle{Annu. Rev. Nucl. Part. Sci.}
\bvolume{45},
\bfpage{591}--\blpage{634}
(\byear{1995}).
\doiurl{10.1146/annurev.ns.45.120195.003111}
\end{barticle}
\endbibitem

\bibitem{Trzcinska2001}
\begin{barticle}
\bauthor{\bsnm{Trzci{\'n}ska}, \binits{A.}},
\bauthor{\bsnm{Jastrz{\c e}bski}, \binits{J.}},
\bauthor{\bsnm{Lubi{\'n}ski}, \binits{P.}},
\bauthor{\bsnm{Hartmann}, \binits{F.J.}},
\bauthor{\bsnm{Schmidt}, \binits{R.}},
\bauthor{\bparticle{von} \bsnm{Egidy}, \binits{T.}},
\bauthor{\bsnm{K{\l}os}, \binits{B.}}:
\batitle{{Neutron density distributions deduced from antiprotonic atoms}}.
\bjtitle{Phys. Rev. Lett.}
\bvolume{87},
\bfpage{82501}
(\byear{2001}).
\doiurl{10.1103/PhysRevLett.87.082501}
\end{barticle}
\endbibitem

\bibitem{Angeli2013}
\begin{barticle}
\bauthor{\bsnm{Angeli}, \binits{I.}},
\bauthor{\bsnm{Marinova}, \binits{K.P.}}:
\batitle{{Table of experimental nuclear ground state charge radii: An update}}.
\bjtitle{At. Data Nucl. Data Tables}
\bvolume{99},
\bfpage{69}--\blpage{95}
(\byear{2013}).
\doiurl{10.1016/j.adt.2011.12.006}
\end{barticle}
\endbibitem

\bibitem{Brown2000}
\begin{barticle}
\bauthor{\bsnm{Brown}, \binits{B.A.}}:
\batitle{{Neutron radii in nuclei and the neutron equation of state}}.
\bjtitle{Phys. Rev. Lett.}
\bvolume{85},
\bfpage{5296}--\blpage{5299}
(\byear{2000}).
\doiurl{10.1103/PhysRevLett.85.5296}
\end{barticle}
\endbibitem

\bibitem{Centelles2010}
\begin{barticle}
\bauthor{\bsnm{Centelles}, \binits{M.}},
\bauthor{\bsnm{Roca-Maza}, \binits{X.}},
\bauthor{\bsnm{Vi{\~n}as}, \binits{X.}},
\bauthor{\bsnm{Warda}, \binits{M.}}:
\batitle{{Origin of the neutron skin thickness of $^{208}$Pb in nuclear
  mean-field models}}.
\bjtitle{Phys. Rev. C}
\bvolume{82},
\bfpage{054314}
(\byear{2010}).
\doiurl{10.1103/PhysRevC.82.054314}
\end{barticle}
\endbibitem

\bibitem{Warda2010}
\begin{barticle}
\bauthor{\bsnm{Warda}, \binits{M.}},
\bauthor{\bsnm{Vi{\~n}as}, \binits{X.}},
\bauthor{\bsnm{Roca-Maza}, \binits{X.}},
\bauthor{\bsnm{Centelles}, \binits{M.}}:
\batitle{{Analysis of bulk and surface contributions in the neutron skin of
  nuclei}}.
\bjtitle{Phys. Rev. C}
\bvolume{81},
\bfpage{054309}
(\byear{2010}).
\doiurl{10.1103/PhysRevC.81.054309}
\end{barticle}
\endbibitem

\bibitem{Tarbert2014}
\begin{barticle}
\bauthor{\bsnm{Tarbert}, \binits{C.M.}},
\bauthor{\bsnm{Watts}, \binits{D.P.}},
\bauthor{\bsnm{Glazier}, \binits{D.I.}}, \betal:
\batitle{{{Neutron Skin of $^{208}$Pb from Coherent Pion Photoproduction}}}.
\bjtitle{Phys. Rev. Lett.}
\bvolume{112},
\bfpage{242502}
(\byear{2014}).
\doiurl{10.1103/PhysRevLett.112.242502}
\end{barticle}
\endbibitem

\bibitem{Adhikari2021}
\begin{barticle}
\bauthor{\bsnm{Adhikari}, \binits{D.}},
\bauthor{\bsnm{Albataineh}, \binits{H.}},
\bauthor{\bsnm{Androic}, \binits{D.}}, \betal:
\batitle{{Accurate Determination of the Neutron Skin Thickness of $^{208}$Pb
  through Parity-Violation in Electron Scattering}}.
\bjtitle{Phys. Rev. Lett.}
\bvolume{126},
\bfpage{172502}
(\byear{2021}).
\doiurl{10.1103/PhysRevLett.126.172502}
\end{barticle}
\endbibitem

\bibitem{Dobaczewski1996}
\begin{barticle}
\bauthor{\bsnm{Dobaczewski}, \binits{J.}},
\bauthor{\bsnm{Nazarewicz}, \binits{W.}},
\bauthor{\bsnm{Werner}, \binits{T.R.}}:
\batitle{{Neutron radii and skins in the Hartree-Fock-Bogoliubov
  calculations}}.
\bjtitle{Z. Phys. A}
\bvolume{354},
\bfpage{27}--\blpage{35}
(\byear{1996}).
\doiurl{10.1007/s002180050009}
\end{barticle}
\endbibitem

\bibitem{Horowitz2001}
\begin{barticle}
\bauthor{\bsnm{Horowitz}, \binits{C.J.}},
\bauthor{\bsnm{Piekarewicz}, \binits{J.}}:
\batitle{{Neutron star structure and the neutron radius of $^{208}$Pb}}.
\bjtitle{Phys. Rev. Lett.}
\bvolume{86},
\bfpage{5647}
(\byear{2001}).
\doiurl{10.1103/PhysRevLett.86.5647}
\end{barticle}
\endbibitem

\bibitem{Steiner2005}
\begin{barticle}
\bauthor{\bsnm{Steiner}, \binits{A.W.}},
\bauthor{\bsnm{Prakash}, \binits{M.}},
\bauthor{\bsnm{Lattimer}, \binits{J.M.}},
\bauthor{\bsnm{Ellis}, \binits{P.J.}}:
\batitle{{Isospin asymmetry in nuclei and neutron stars}}.
\bjtitle{Phys. Rep.}
\bvolume{411},
\bfpage{325}--\blpage{375}
(\byear{2005}).
\doiurl{10.1016/j.physrep.2005.02.004}
\end{barticle}
\endbibitem

\bibitem{Fang2011}
\begin{barticle}
\bauthor{\bsnm{Fang}, \binits{D.Q.}},
\bauthor{\bsnm{Ma}, \binits{Y.G.}},
\bauthor{\bsnm{Cai}, \binits{X.Z.}},
\bauthor{\bsnm{Tian}, \binits{W.D.}},
\bauthor{\bsnm{Wang}, \binits{H.W.}}:
\batitle{{Effects of neutron skin thickness in peripheral nuclear reactions}}.
\bjtitle{Chin. Phys. Lett.}
\bvolume{28},
\bfpage{10}--\blpage{13}
(\byear{2011}).
\doiurl{10.1088/0256-307X/28/10/102102}
\end{barticle}
\endbibitem

\bibitem{Fang2010}
\begin{barticle}
\bauthor{\bsnm{Fang}, \binits{D.Q.}},
\bauthor{\bsnm{Ma}, \binits{Y.G.}},
\bauthor{\bsnm{Cai}, \binits{X.Z.}},
\bauthor{\bsnm{Tian}, \binits{W.D.}},
\bauthor{\bsnm{Wang}, \binits{H.W.}}:
\batitle{{Neutron removal cross section as a measure of neutron skin}}.
\bjtitle{Phys. Rev. C}
\bvolume{81},
\bfpage{047603}
(\byear{2010}).
\doiurl{10.1103/PhysRevC.81.047603}
\end{barticle}
\endbibitem

\bibitem{Yan2019}
\begin{barticle}
\bauthor{\bsnm{Yan}, \binits{T.-Z.}},
\bauthor{\bsnm{Li}, \binits{S.}}:
\batitle{{Impact parameter dependence of the yield ratios of light particles as
  a probe of neutron skin}}.
\bjtitle{Nucl. Sci. Tech.}
\bvolume{30},
\bfpage{43}
(\byear{2019}).
\doiurl{10.1007/s41365-019-0572-8}
\end{barticle}
\endbibitem

\bibitem{Aumann2017}
\begin{barticle}
\bauthor{\bsnm{Aumann}, \binits{T.}},
\bauthor{\bsnm{Bertulani}, \binits{C.A.}},
\bauthor{\bsnm{Schindler}, \binits{F.}},
\bauthor{\bsnm{Typel}, \binits{S.}}:
\batitle{{Peeling Off Neutron Skins from Neutron-Rich Nuclei: Constraints on
  the Symmetry Energy from Neutron-Removal Cross Sections}}.
\bjtitle{Phys. Rev. Lett.}
\bvolume{119},
\bfpage{262501}
(\byear{2017}).
\doiurl{10.1103/PhysRevLett.119.262501}
\end{barticle}
\endbibitem

\bibitem{Bertulani2019}
\begin{barticle}
\bauthor{\bsnm{Bertulani}, \binits{C.A.}},
\bauthor{\bsnm{Valencia}, \binits{J.}}:
\batitle{{Neutron skins as laboratory constraints on properties of neutron
  stars and on what we can learn from heavy ion fragmentation reactions}}.
\bjtitle{Phys. Rev. C}
\bvolume{100},
\bfpage{015802}
(\byear{2019}).
\doiurl{10.1103/PhysRevC.100.015802}
\end{barticle}
\endbibitem

\bibitem{De2017}
\begin{barticle}
\bauthor{\bsnm{De}, \binits{S.}}:
\batitle{{The effect of neutron skin on inclusive prompt photon production in
  Pb + Pb collisions at Large Hadron Collider energies}}.
\bjtitle{J. Phys. G: Nucl. Part. Phys.}
\bvolume{44},
\bfpage{045104}
(\byear{2017}).
\doiurl{10.1088/1361-6471/aa5689}
\end{barticle}
\endbibitem

\bibitem{Paukkunen2015}
\begin{barticle}
\bauthor{\bsnm{Paukkunen}, \binits{H.}}:
\batitle{{Neutron skin and centrality classification in high-energy heavy-ion
  collisions at the LHC}}.
\bjtitle{Phys. Lett. B}
\bvolume{745},
\bfpage{73}--\blpage{78}
(\byear{2015}).
\doiurl{10.1016/j.physletb.2015.04.037}
\end{barticle}
\endbibitem

\bibitem{Alvioli2019}
\begin{barticle}
\bauthor{\bsnm{Alvioli}, \binits{M.}},
\bauthor{\bsnm{Strikman}, \binits{M.}}:
\batitle{{Spin-isospin correlated configurations in complex nuclei and neutron
  skin effect in W$^\pm$ production in high-energy proton-lead collisions}}.
\bjtitle{Phys. Rev. C}
\bvolume{100},
\bfpage{024912}
(\byear{2019}).
\doiurl{10.1103/PhysRevC.100.024912}
\end{barticle}
\endbibitem

\bibitem{Li2020}
\begin{barticle}
\bauthor{\bsnm{Li}, \binits{H.}},
\bauthor{\bsnm{Xu}, \binits{H.-j.}},
\bauthor{\bsnm{Zhou}, \binits{Y.}},
\bauthor{\bsnm{Wang}, \binits{X.}},
\bauthor{\bsnm{Zhao}, \binits{J.}},
\bauthor{\bsnm{Chen}, \binits{L.-W.}},
\bauthor{\bsnm{Wang}, \binits{F.}}:
\batitle{Probing the neutron skin with ultrarelativistic isobaric collisions}.
\bjtitle{Phys. Rev. Lett.}
\bvolume{125},
\bfpage{222301}
(\byear{2020}).
\doiurl{10.1103/physrevlett.125.222301}
\end{barticle}
\endbibitem

\bibitem{Pshenichnov2021properties}
\begin{barticle}
\bauthor{\bsnm{Pshenichnov}, \binits{I.A.}},
\bauthor{\bsnm{Kozyrev}, \binits{N.A.}},
\bauthor{\bsnm{Nepeivoda}, \binits{R.S.}},
\bauthor{\bsnm{Svetlichnyi}, \binits{A.O.}},
\bauthor{\bsnm{Dmitrieva}, \binits{N.A.}}:
\batitle{Properties of spectator matter in nuclear collisions at {NICA}}.
\bjtitle{Phys. Part. Nucl.}
\bvolume{52},
\bfpage{591}--\blpage{597}
(\byear{2021}).
\doiurl{10.1134/S1063779621040493}
\end{barticle}
\endbibitem

\bibitem{Dmitrieva2021}
\begin{barticle}
\bauthor{\bsnm{Dmitrieva}, \binits{U.}},
\bauthor{\bsnm{Kozyrev}, \binits{N.}},
\bauthor{\bsnm{Svetlichnyi}, \binits{A.}},
\bauthor{\bsnm{Pshenichnov}, \binits{I.}}:
\batitle{{Spectator nucleons in most central collisions of heavy nuclei at
  NICA}}.
\bjtitle{AIP Conf. Proc.}
\bvolume{2377},
\bfpage{030005}
(\byear{2021}).
\doiurl{10.1063/5.0063284}
\end{barticle}
\endbibitem

\bibitem{Kozyrev2021}
\begin{bchapter}
\bauthor{\bsnm{Kozyrev}, \binits{N.}},
\bauthor{\bsnm{Svetlichnyi}, \binits{A.}},
\bauthor{\bsnm{Nepeivoda}, \binits{R.}},
\bauthor{\bsnm{Pshenichnov}, \binits{I.A.}}:
\bctitle{{Spectator nucleons in ultracentral $^{208}$Pb–$^{208}$Pb collisions
  as a probe of nuclear periphery}}.
In: \bbtitle{Proceedings of The Ninth Annual Conference on Large Hadron
  Collider Physics — PoS(LHCP2021)},
p. \bfpage{223}.
\bpublisher{Sissa Medialab},
\blocation{Trieste, Italy}
(\byear{2021}).
\doiurl{10.22323/1.397.0223}.
\burl{https://pos.sissa.it/397/223}
\end{bchapter}
\endbibitem

\bibitem{Pshenichnov2022}
\begin{barticle}
\bauthor{\bsnm{Pshenichnov}, \binits{I.A.}},
\bauthor{\bsnm{Kozyrev}, \binits{N.A.}},
\bauthor{\bsnm{Svetlichnyi}, \binits{A.O.}},
\bauthor{\bsnm{Dmitrieva}, \binits{U.A.}}:
\batitle{{What One Can Learn by Studying Spectator Remnants in Central
  Nucleus–Nucleus Collisions?}}
\bjtitle{Phys. Part. and Nucl.}
\bvolume{53},
\bfpage{335}--\blpage{341}
(\byear{2022}).
\doiurl{10.1134/S1063779622020691}
\end{barticle}
\endbibitem

\bibitem{Liu2022}
\begin{botherref}
\oauthor{\bsnm{Liu}, \binits{L.-M.}},
\oauthor{\bsnm{Zhang}, \binits{C.-J.}},
\oauthor{\bsnm{Zhou}, \binits{J.}},
\oauthor{\bsnm{Xu}, \binits{J.}},
\oauthor{\bsnm{Jia}, \binits{J.}},
\oauthor{\bsnm{Peng}, \binits{G.-X.}}:
{Probing neutron-skin thickness with free spectator neutrons in ultracentral
  high-energy isobaric collisions}
(2022)
{\href{https://arxiv.org/abs/2203.09924}{{arXiv:2203.09924}}}
\end{botherref}
\endbibitem

\bibitem{svetlichnyi2021using}
\begin{barticle}
\bauthor{\bsnm{Svetlichnyi}, \binits{A.}},
\bauthor{\bsnm{Nepeyvoda}, \binits{R.}},
\bauthor{\bsnm{Pshenichnov}, \binits{I.}}:
\batitle{Using spectator matter for centrality determination in nucleus-nucleus
  collisions}.
\bjtitle{Particles}
\bvolume{4},
\bfpage{227}--\blpage{235}
(\byear{2021}).
\doiurl{10.3390/particles4020021}
\end{barticle}
\endbibitem

\bibitem{svetlichnyi2020formation}
\begin{barticle}
\bauthor{\bsnm{Svetlichnyi}, \binits{A.}},
\bauthor{\bsnm{Pshenichnov}, \binits{I.}}:
\batitle{Formation of free and bound spectator nucleons in hadronic
  interactions between relativistic nuclei}.
\bjtitle{Bull. Russ. Acad. Sci. Phys.}
\bvolume{84},
\bfpage{911}--\blpage{916}
(\byear{2020}).
\doiurl{10.3103/S1062873820080110}
\end{barticle}
\endbibitem

\bibitem{nepeivoda2021dependence}
\begin{botherref}
\oauthor{\bsnm{Nepeivoda}, \binits{R.S.}},
\oauthor{\bsnm{Svetlichnyi}, \binits{A.O.}}:
Dependence of n/p-ratio in spectator matter on the energy and mass of colliding
  nuclei.
Memoirs of the Faculty of Physics
(1),
2110302
(2021)
\end{botherref}
\endbibitem

\bibitem{Nepeivoda2022}
\begin{barticle}
\bauthor{\bsnm{Nepeivoda}, \binits{R.}},
\bauthor{\bsnm{Svetlichnyi}, \binits{A.}},
\bauthor{\bsnm{Kozyrev}, \binits{N.}},
\bauthor{\bsnm{Pshenichnov}, \binits{I.}}:
\batitle{{Pre-Equilibrium Clustering in Production of Spectator Fragments in
  Collisions of Relativistic Nuclei}}.
\bjtitle{Particles}
\bvolume{5},
\bfpage{40}--\blpage{51}
(\byear{2022}).
\doiurl{10.3390/particles5010004}
\end{barticle}
\endbibitem

\bibitem{Loizides2018}
\begin{barticle}
\bauthor{\bsnm{Loizides}, \binits{C.}},
\bauthor{\bsnm{Kamin}, \binits{J.}},
\bauthor{\bsnm{D'Enterria}, \binits{D.}}:
\batitle{{Improved Monte Carlo Glauber predictions at present and future
  nuclear colliders}}.
\bjtitle{Phys. Rev. C}
\bvolume{97},
\bfpage{054910}
(\byear{2018}).
\doiurl{10.1103/PhysRevC.97.054910}
\end{barticle}
\endbibitem

\bibitem{Ericson1960}
\begin{barticle}
\bauthor{\bsnm{Ericson}, \binits{T.}}:
\batitle{{The statistical model and nuclear level densities}}.
\bjtitle{Adv. Phys.}
\bvolume{9},
\bfpage{425}--\blpage{511}
(\byear{1960}).
\doiurl{10.1080/00018736000101239}
\end{barticle}
\endbibitem

\bibitem{Weisskopf1940}
\begin{barticle}
\bauthor{\bsnm{Weisskopf}, \binits{V.F.}},
\bauthor{\bsnm{Ewing}, \binits{D.H.}}:
\batitle{{On the yield of nuclear reactions with heavy elements}}.
\bjtitle{Phys. Rev.}
\bvolume{57},
\bfpage{472}--\blpage{485}
(\byear{1940}).
\doiurl{10.1103/PhysRev.57.472}
\end{barticle}
\endbibitem

\bibitem{Fermi1950}
\begin{barticle}
\bauthor{\bsnm{Fermi}, \binits{E.}}:
\batitle{{High Energy Nuclear Events}}.
\bjtitle{Prog. Theor. Phys.}
\bvolume{5},
\bfpage{570}--\blpage{583}
(\byear{1950}).
\doiurl{10.1143/ptp/5.4.570}
\end{barticle}
\endbibitem

\bibitem{Bondorf1995}
\begin{barticle}
\bauthor{\bsnm{Bondorf}, \binits{J.P.}},
\bauthor{\bsnm{Botvina}, \binits{A.S.}},
\bauthor{\bsnm{Iljinov}, \binits{A.S.}},
\bauthor{\bsnm{Mishustin}, \binits{I.N.}},
\bauthor{\bsnm{Sneppen}, \binits{K.}}:
\batitle{{Statistical multifragmentation of nuclei}}.
\bjtitle{Phys. Rep.}
\bvolume{257},
\bfpage{133}--\blpage{221}
(\byear{1995}).
\doiurl{10.1016/0370-1573(94)00097-M}
\end{barticle}
\endbibitem

\bibitem{Agostinelli2003}
\begin{barticle}
\bauthor{\bsnm{Agostinelli}, \binits{S.}},
\bauthor{\bsnm{Allison}, \binits{J.}},
\bauthor{\bsnm{Amako}, \binits{K.}},
\bauthor{\bsnm{Apostolakis}, \binits{J.}}, \betal:
\batitle{{Geant4 -- a simulation toolkit}}.
\bjtitle{Nucl. Inst. Meth. A}
\bvolume{506},
\bfpage{250}--\blpage{303}
(\byear{2003}).
\doiurl{10.1016/S0168-9002(03)01368-8}
\end{barticle}
\endbibitem

\bibitem{Loizides2016}
\begin{barticle}
\bauthor{\bsnm{Loizides}, \binits{C.}}:
\batitle{Glauber modeling of high-energy nuclear collisions at the subnucleon
  level}.
\bjtitle{Phys. Rev. C}
\bvolume{94},
\bfpage{024914}
(\byear{2016}).
\doiurl{10.1103/PhysRevC.94.024914}
\end{barticle}
\endbibitem

\bibitem{Gaimard1991}
\begin{barticle}
\bauthor{\bsnm{Gaimard}, \binits{J.J.}},
\bauthor{\bsnm{Schmidt}, \binits{K.H.}}:
\batitle{A reexamination of the abrasion-ablation model for the description of
  the nuclear fragmentation reaction}.
\bjtitle{Nucl. Phys. A}
\bvolume{531},
\bfpage{709}--\blpage{745}
(\byear{1991}).
\doiurl{10.1016/0375-9474(91)90748-U}
\end{barticle}
\endbibitem

\bibitem{Scheidenberger2004}
\begin{barticle}
\bauthor{\bsnm{Scheidenberger}, \binits{C.}},
\bauthor{\bsnm{Pshenichnov}, \binits{I.A.}},
\bauthor{\bsnm{S{\"u}mmerer}, \binits{K.}}, \betal:
\batitle{{Charge-changing interactions of ultrarelativistic Pb nuclei}}.
\bjtitle{Phys. Rev. C}
\bvolume{70},
\bfpage{014902}
(\byear{2004}).
\doiurl{10.1103/PhysRevC.70.014902}
\end{barticle}
\endbibitem

\bibitem{CollaborationALICE2020}
\begin{botherref}
\oauthor{\bsnm{Chinellato}, \binits{D.D.}}, et al.:
{Data-driven model for the emission of spectator nucleons as a function of
  centrality in Pb-Pb collisions at LHC energies, ALICE-PUBLIC-2020-001}
(2020).
\url{http://cds.cern.ch/record/2712412}
\end{botherref}
\endbibitem

\bibitem{Botvina2022}
\begin{barticle}
\bauthor{\bsnm{Botvina}, \binits{A.S.}},
\bauthor{\bsnm{Buyukcizmeci}, \binits{N.}},
\bauthor{\bsnm{Bleicher}, \binits{M.}}:
\batitle{Evolution of the statistical disintegration of finite nuclei toward
  high energy}.
\bjtitle{Phys. Rev. C}
\bvolume{106},
\bfpage{014607}
(\byear{2022}).
\doiurl{10.1103/PhysRevC.106.014607}
\end{barticle}
\endbibitem

\bibitem{Prim1957}
\begin{botherref}
\oauthor{\bsnm{Prim}, \binits{R.C.}}:
{Bell Syst. Tech. J.}
\textbf{36},
1389--1401
(1957).
\doiurl{10.1002/j.1538-7305.1957.tb01515.x}
\end{botherref}
\endbibitem

\bibitem{Viola2004}
\begin{barticle}
\bauthor{\bsnm{Viola}, \binits{V.E.}},
\bauthor{\bsnm{Kwiatkowski}, \binits{K.}},
\bauthor{\bsnm{Natowitz}, \binits{J.B.}},
\bauthor{\bsnm{Yennello}, \binits{S.J.}}:
\batitle{{Breakup densities of hot nuclei}}.
\bjtitle{Phys. Rev. Lett.}
\bvolume{93},
\bfpage{132701}
(\byear{2004}).
\doiurl{10.1103/PhysRevLett.93.132701}
\end{barticle}
\endbibitem

\bibitem{De2006}
\begin{barticle}
\bauthor{\bsnm{De}, \binits{J.N.}},
\bauthor{\bsnm{Samaddar}, \binits{S.K.}},
\bauthor{\bsnm{Vi{\~n}as}, \binits{X.}},
\bauthor{\bsnm{Centelles}, \binits{M.}}:
\batitle{{Nuclear expansion with excitation}}.
\bjtitle{Phys. Lett. B}
\bvolume{638},
\bfpage{160}--\blpage{165}
(\byear{2006}).
\doiurl{10.1016/j.physletb.2006.05.046}
\end{barticle}
\endbibitem

\bibitem{Allison2016}
\begin{barticle}
\bauthor{\bsnm{Allison}, \binits{J.}},
\bauthor{\bsnm{Amako}, \binits{K.}},
\bauthor{\bsnm{Apostolakis}, \binits{J.}},
\bauthor{\bsnm{Arce}, \binits{P.}},
\bauthor{\bsnm{Asai}, \binits{M.}},
\bauthor{\bsnm{Aso}, \binits{T.}},
\bauthor{\bsnm{Bagli}, \binits{E.}},
\bauthor{\bsnm{Bagulya}, \binits{A.}},
\bauthor{\bsnm{Banerjee}, \binits{S.}},
\bauthor{\bsnm{Barrand}, \binits{G.}}, \betal:
\batitle{Recent developments in {Geant4}}.
\bjtitle{Nucl. Instrum. Methods A}
\bvolume{835},
\bfpage{186}--\blpage{225}
(\byear{2016}).
\doiurl{10.1016/j.nima.2016.06.125}
\end{barticle}
\endbibitem

\bibitem{Weisskopf1937}
\begin{barticle}
\bauthor{\bsnm{Weisskopf}, \binits{V.}}:
\batitle{{Statistics and Nuclear Reactions}}.
\bjtitle{Phys. Rev.}
\bvolume{52},
\bfpage{295}--\blpage{303}
(\byear{1937}).
\doiurl{10.1103/PhysRev.52.295}
\end{barticle}
\endbibitem

\bibitem{Dmitrieva2018}
\begin{barticle}
\bauthor{\bsnm{Dmitrieva}, \binits{U.}},
\bauthor{\bsnm{Pshenichnov}, \binits{I.}}:
\batitle{{On the performance of Zero Degree Calorimeters in detecting
  multinucleon events}}.
\bjtitle{Nucl. Instrum. Methods Phys. Res., Sect. A}
\bvolume{906},
\bfpage{114}--\blpage{119}
(\byear{2018}).
\doiurl{10.1016/j.nima.2018.07.072}
\end{barticle}
\endbibitem

\bibitem{Alvioli2012}
\begin{barticle}
\bauthor{\bsnm{Alvioli}, \binits{M.}},
\bauthor{\bsnm{Holopainen}, \binits{H.}},
\bauthor{\bsnm{Eskola}, \binits{K.J.}},
\bauthor{\bsnm{Strikman}, \binits{M.}}:
\batitle{Initial-state anisotropies and their uncertainties in
  ultrarelativistic heavy-ion collisions from the monte carlo glauber model}.
\bjtitle{Phys. Rev. C}
\bvolume{85},
\bfpage{034902}
(\byear{2012}).
\doiurl{10.1103/PhysRevC.85.034902}
\end{barticle}
\endbibitem

\bibitem{Alvioli2011}
\begin{barticle}
\bauthor{\bsnm{Alvioli}, \binits{M.}},
\bauthor{\bsnm{Strikman}, \binits{M.}}:
\batitle{Beam fragmentation in heavy ion collisions with realistically
  correlated nuclear configurations}.
\bjtitle{Phys. Rev. C}
\bvolume{83},
\bfpage{044905}
(\byear{2011}).
\doiurl{10.1103/PhysRevC.83.044905}
\end{barticle}
\endbibitem

\bibitem{Alvioli2009}
\begin{barticle}
\bauthor{\bsnm{Alvioli}, \binits{M.}},
\bauthor{\bsnm{Drescher}, \binits{H.J.}},
\bauthor{\bsnm{Strikman}, \binits{M.}}:
\batitle{{A Monte Carlo generator of nucleon configurations in complex nuclei
  including nucleon–nucleon correlations}}.
\bjtitle{Physics Letters B}
\bvolume{680},
\bfpage{225}--\blpage{230}
(\byear{2009}).
\doiurl{10.1016/j.physletb.2009.08.067}
\end{barticle}
\endbibitem

\bibitem{Abelev2012n}
\begin{barticle}
\bauthor{\bsnm{Abelev}, \binits{B.}},
\bauthor{\bsnm{Adam}, \binits{J.}},
\bauthor{\bsnm{Adamov{\'a}}, \binits{D.}}, \betal:
\batitle{{Measurement of the Cross Section for Electromagnetic Dissociation
  with Neutron Emission in Pb--Pb collisions at $\sqrt{s_{\mathrm NN}}=2.76$
  TeV}}.
\bjtitle{Phys. Rev. Lett.}
\bvolume{109},
\bfpage{252302}
(\byear{2012}).
\doiurl{10.1103/PhysRevLett.109.252302}
\end{barticle}
\endbibitem

\end{thebibliography}

\end{document}